\RequirePackage{fix-cm}
\RequirePackage{rotating}

\documentclass[smallextended]{svjour3}       
\smartqed  
\usepackage[lflt]{floatflt} 
\usepackage{graphicx}

\usepackage[dvipsnames]{xcolor} 
\usepackage{soul} 

\usepackage[title]{appendix} 

\usepackage{booktabs} 
\usepackage{soulutf8} 
\usepackage{rotating} 
\usepackage[utf8]{inputenc} 
\usepackage{multirow} 
\usepackage{colortbl} 
\usepackage{framed} 

\definecolor{lightgray}{gray}{0.9}
\definecolor{mediumgray}{gray}{0.7}
\definecolor{darkgray}{gray}{0.5}

\begin{document}

\title{On (Mis)Perceptions of Testing Effectiveness: An Empirical Study\thanks{This research was funded by Spanish Ministry of Science, Innovation and Universities research grant PGC2018-097265-B-I00, the Regional Government of Madrid, under the FORTE-CM project (S2018/TCS-4314) and the Spanish Ministry of Economy and Business, under the  MADRID project (TIN2017-88557-R).}
}


\author{Sira Vegas         \and
        Patricia Riofrío \and
        Esperanza Marcos \and
        Natalia Juristo 
}


\institute{Sira Vegas \at
              Universidad Politécnica de Madrid, Madrid, Spain. \\
              \email{svegas@fi.upm.es}           
           \and
           Patricia Riofrío \at
              Universidad Politécnica de Madrid, Madrid, Spain. \\
           \and
           Esperanza Marcos \at
              Universidad Rey Juan Carlos, Madrid, Spain. \\
           \and
           Natalia Juristo \at
              Universidad Politécnica de Madrid, Madrid, Spain. \\
}


\maketitle

\begin{abstract}
A recurring problem in software development is incorrect decision making on the techniques, methods and tools to be used. Mostly, these decisions are based on developers' perceptions about them. A factor influencing people's perceptions is past experience, but it is not the only one. In this research, we aim to discover how well the perceptions of the defect detection effectiveness of different techniques match their real effectiveness in the absence of prior experience. To do this, we conduct an empirical study plus a replication. During the original study, we conduct a controlled experiment with students applying two testing techniques and a code review technique. At the end of the experiment, they take a survey to find out which technique they perceive to be most effective. The results show that participants' perceptions are wrong and that this mismatch is costly in terms of quality. In order to gain further insight into the results, we replicate the controlled experiment and extend the survey to include questions about participants' opinions on the techniques and programs. The results of the replicated study confirm the findings of the original study and suggest that participants' perceptions might be based not on their opinions about complexity or preferences for techniques but on how well they think that they have applied the techniques.
\end{abstract}

\section{Introduction}

An increasingly more popular practice nowadays is for software development companies to let developers choose their own technological environment. This means that different developers may use different productivity tools (programming language, IDE, etc.). However, software engineering (SE) is a human-intensive discipline where wrong decisions can potentially compromise the quality of the resulting software.

In SE, decisions on which methods, techniques and tools to use in software development are typically based on developers' perceptions and/or opinions rather than evidence, as suggested by Dybå \textit{et al.} \cite{DKJ:05} and Zelkowitz \textit{et al.} \cite{ZWB:03}. However, empirical evidence might not be available, as certain methods, techniques or tools may not have been studied within a particular setting or even at all. Alternatively, developers may simply not be acquainted with such studies, according to Vegas \& Basili \cite{VB:05}. On this ground, it is important to discover how well developers perceptions  (beliefs) match reality and, if they do not, find out what is behind this mismatch, as noted by Devanbu \textit{et al.}\cite{DZB:16}. 

According to Psychology, experience plays a role in people's perceptions. This has also been observed by Devanbu \textit{et al.} \cite{DZB:16} in SE. However, this research sets out to discover how well matched perceptions are with reality in the absence of previous experience in the technology being used. This makes sense for several reasons: 1) experience is not the only factor affecting developers' perceptions; 2) development teams are usually composed of a mix of people with and without experience; and 3) it is not clear what type of experience influences perceptions. For example, Dieste \textit{et al.} \cite{DAUTTFOJ:17} conclude that academic rather than professional experience could be affecting the external quality of the code generated by developers when applying Test-Driven Development.
  
We aim to study whether perceptions about the effectiveness of three defect detection techniques match reality, and if not, what is behind these perceptions. To the best of our knowledge, this is the first paper to empirically assess this issue. 

To this end, we conducted an empirical study plus a replication with students. During the original study we measured (as part of a controlled experiment) the effectiveness of two testing techniques and one code review technique when applied by the participants. We then checked the perceived most effective technique (gathered by means of a survey) against the real one. Additionally, we analysed the cost of the mismatch between perceptions and reality in terms of loss of effectiveness. Major findings include:

\begin{itemize}

\item Different people perceive different techniques to be more effective. No one technique is perceived as being more effective than the others.

\item The perceptions of 50\% of participants (11 out of 23) are wrong.

\item Wrong perception of techniques can reduce effectiveness 31pp (percentage points) on average.

\end{itemize}

These findings led us to extend the goal of the study in a replication to investigate what could be behind participants' perceptions. To do this, we examined their opinions on the techniques they applied and the programs they tested in a replication of the controlled experiment. Major findings include:

\begin{itemize}

\item The results of the replication confirm the findings of the original study.

\item Participants think that technique effectiveness depends exclusively on their performance and not on possible weaknesses of the technique itself.

\item The opinions about technique complexity and preferences for techniques do not seem to play a role in perceived effectiveness. 

\end{itemize}

These results are useful for developers and researchers. They suggest: 

\begin{itemize}

\item Developers should become aware of the limitations of their judgement.

\item Tools should be designed that provide feedback to developers on how effective techniques are.

\item The best combination of techniques to apply should be determined that is at the same time easily applicable and effective.

\item Instruments should be developed to make empirical results available to developers. 

\end{itemize}

The material associated to the studies presented here can be found at https://github.com/GRISE-UPM/Misperceptions.

The article is organised as follows. Section~\ref{sec:methodology_it1} describes the original study. Section~\ref{sec:threats_it1} presents its validity threats. Section~\ref{sec:results_it1} discusses the results. Section~\ref{sec:methodology_it2} describes the replicated study based on the modifications made to the original study. Section~\ref{sec:threats_it2} presents its validity threats. Section~\ref{sec:results_it2} reports the results of this replicated study. Section~\ref{sec:discussion} discusses our findings and their implications. Section~\ref{sec:related_work} shows related work. Finally, Section~\ref{sec:conclusions} outlines the conclusions of this work.

\section{Original Study: Research Questions and Methodology}
\label{sec:methodology_it1}

\subsection{Research Questions}

The main goal of the original study is to assess whether participants' perceptions of their testing effectiveness using different techniques are good predictors of real testing effectiveness. This goal has been translated into the following research question:

RQ1: Should participants' perceptions be used as predictors of testing effectiveness?

This question was further decomposed into:

\begin{itemize}

    \item \textit{RQ1.1: What are participants' perceptions of their testing effectiveness?} 
    
    We want to know if participants perceive a certain technique as most effective than the others.

    \item \textit{RQ1.2: Do participants' perceptions predict their testing effectiveness?} 
    
    We want to assess if the technique each participant perceives as most effective is the most effective for him/her.

    \item \textit{RQ1.3: Do participants find a similar amount of defects for all techniques?}
   	
   	Choosing the most effective technique can be difficult if participants find a similar amount of defects for two or all three techniques.

    \item \textit{RQ1.4: What is the cost of any mismatch?} 
    
    We want to know if the cost of not correctly perceiving the most effective technique is negligible and depends on the technique perceived as most effective.

    \item \textit{RQ1.5: What is expected project loss?} 

	Taking into consideration that some participants will correctly perceive their most effective technique (mismatch cost 0), and others will not (mismatch cost greater than 0), we calculate the overall cost of (mis)match for all participants in the empirical study and check if it depends on the technique perceived as most effective. 

\end{itemize}

\subsection{Study Context and Ethics}

We conducted a controlled experiment where each participant applies three defect detection techniques (two testing techniques and one code review technique) on three different programs. For testing techniques, participants report the generated test cases, later run a set of test cases that we have generated (instead of the ones they created), and report the failures found\footnote{This has been done for learning purposes, as we have noticed that students sometimes do not report failures that are exercised by test cases. Since this is a learning goal of the course, not relevant for the study, we measure it separately, and do not use it here.}. For code reading they report the identified faults. At the end of the controlled experiment, each participant completes a questionnaire containing a question related to his/her perceptions of the effectiveness of the techniques applied. The course is graded based on their technique application performance (this guarantees a thorough application of the techniques).

The study is embedded in an elective 6 credits Software Verification and Validation course. The regular assessment (when the experiment does not take place) is as follows: students are asked to write a specification for a program that can be coded in about 8 hours. Specifications are later interchanged so that each student codes a different program from the one (s)he proposed. Later, students are asked to perform individually (in successive weeks) code reading, and white-box testing on the code they wrote. At this point, each student delivers the code to the person who wrote the specification, so that each student performs black-box testing on the program (s)he proposed. Note that this scenario requires more effort from the student (as (s)he is asked to write first a specification and then code a program, and these tasks do not take place when the study is run). In other words, the students workload during the experiment is smaller than the workload of the regular course assessment. The only activity that takes place during the experiment that is not part of the regular course is answering the questionnaire, which can be done in less than 15 minutes. Although the study causes changes in the workflow of the course, its learning goals are not altered.

All tasks required by the study, with the exception of completing the questionnaire, take place during the slots assigned to the course. Therefore, there is no additional effort for the students but attending lectures (which is mandatory in any case).

Note  that the students are allowed to withdraw from the controlled experiment, but this would affect their score in the course. But this also happens when the experiment is not run. If a student misses one assignment, (s)he would score 0 in that assignment and his/her course score would be affected consequently. However, they are allowed to withdraw from the study without penalty in their score, as the submission of the questionnaire is completely voluntary. No incentives are given to those students who submit the questionnaire.

The fact of submitting the questionnaire implies giving consent for participating in the study. Students are aware this is a voluntary activity aiming for a research, but they can also get feedback. Those students who do not submit the questionnaire, are not considered in the study in any way, as they are not giving consent to use their data. For this reason, they will not be included in the quantitative analysis of the controlled experiment (even though their data is available for scoring purposes).

The study is performed in Spanish, as it is the participants' mother tongue. Its main characteristics are summarised in Table~\ref{tab:experiment_summary}.

\begin{table}[!hbt]
  \caption{Description of the Experiment}
  \label{tab:experiment_summary}
  \begin{tabular}{ll}
    \toprule     

     Aspect & Value \\ \midrule

     Factor & Code evaluation techniques \\ \midrule

	 Treatments & Equivalence partitioning \\	
     & Branch testing \\	
     & Code reading by stepwise abstraction \\	\midrule

     Response variables & Technique effectiveness \\	
     & Perception of effectiveness \\	
     & Mismatch cost \\ \midrule

     Design & 3 period crossover for 3 treatments \\	
     & First training then operation \\	
     & 6 training sessions of 2 hours each \\	
     & 4 hours of individual practice with techniques \\
     & 3 experimental sessions of 4 hours each \\ \midrule

     Experimental Objects & 3 (cmdline, nametbl, ntree) \\	
     & C programming language \\	
     & 2 versions \\	
     & 7 faults injected \\ \midrule

     Participants & 32 \\	
     & Fifth-(final) year undergraduate CS students \\
     & Experienced in programming \\	
     & Experienced with C \\	
     & Trained in SE  \\	
     & Little or none professional experience \\	
     & No testing experience \\ 
     
     \bottomrule  
\end{tabular}
\end{table}

\subsection{Constructs Operationalization}
\label{sec:constructs}

\textit{Code evaluation technique} is an experiment \textbf{factor}, with three treatments (or levels): equivalence partitioning (EP)---see Myers \textit{et al.} \cite{MBS:04}, branch testing (BT)---see Beizer \cite{B:90}, and code reading by stepwise abstraction (CR)---see Linger \cite{L:79}.

The \textbf{response variables} are \textit{technique effectiveness}, \textit{perception of effectiveness} and \textit{mismatch cost}. Technique effectiveness is measured as follows:

\begin{itemize}

    \item For EP and BT, it is the percentage of faults exercised by the set of test cases generated by each participant. In order to measure the response variable, experimenters execute the test cases generated by each participant\footnote{Note that that it is not possible to take measurements on the failures reported by participants, as they do not run their own test cases, but the ones we have given them.}.

    \item For CR, we calculate the percentage of faults correctly reported by each participant (false positives are discarded).

\end{itemize}

Note that dynamic and code review techniques are not directly comparable as they are different technique types (dynamic techniques find failures and code review techniques find faults). However, the comparison is fair, as:

\begin{itemize}

	\item Application time is not taken into account, and participants are given enough time to complete the assigned task.

	\item  All faults injected are detectable by all techniques. Further details about faults, failures and their correspondence is given in Section \ref{sec:experimental_objects}.
	
\end{itemize}

Perception of effectiveness is gathered by means of a questionnaire with one question that reads: \textit{Using which technique did you detect most defects\footnote{During the training, definitions for the terms error, fault and failure are introduced. Additionally, participants are explained that the generic term \textit{defect} is used to refer to both faults and failures indistinctly.}?}

Mismatch cost is measured, for each participant, as the difference between the effectiveness obtained by the participant in the technique (s)he perceives as most effective and  the most effective in reality for him/her. Note that participants neither know the total amount of seeded faults, nor which techniques are best for their colleagues or themselves. This operationalization imitates the reality of testers –who lack such knowledge in real projects. Therefore, the perception is fully subjective (and made in relation with the other two techniques).

 Table~\ref{tab:measuring_mc} shows three examples that show how mismatch cost is measured. Cells in grey background show the technique for which highest effectiveness is observed for the given participant.

\begin{table}[!hbt]
  \caption{Measuring Mismatch Cost}
  \label{tab:measuring_mc}
  \begin{tabular}{ccccccc}
    \toprule     

      & \multicolumn{3}{c}{Observed Effectiveness} & Perceived & &  \\ \cmidrule{2-4}

     Participant &  CR & BT & EP & most effective & Match & Mismatch cost \\
     \midrule      
     P$_{i}$ & 57.14\% & 0\% & \cellcolor{lightgray}83.33\% & CR & No & 26pp \\	
     P$_{j}$ & 42.86\% & 85.71\% & \cellcolor{lightgray}100\% & EP & Yes & 0pp \\	
     P$_{k}$ & 57.14\% & \cellcolor{lightgray}85.71\% & \cellcolor{lightgray}85.71\% & BT & Yes & 0pp \\	
    \bottomrule  
\end{tabular}
\end{table}

The first row shows a situation where the participant perceives as most effective CR, but the most effective for him/her is EP. In this situation, there is a mismatch (misperception) and the associated cost is calculated as the difference in effectiveness between CR and EP. The second row shows a situation where the participant correctly perceives EP as the most effective technique for him/her. In this situation there is a match (correct perception) and therefore, the associated mismatch cost is 0pp. The third row shows a situation where the participant perceives BT as the most effective technique for him/her, and BT and EP are tied as his/her most effective technique. In this situation we consider that there is a match (correct perception), and therefore, the associated mismatch cost is 0pp.

\subsection{Study Design}

Testing techniques are applied by human beings, and no two people are the same. Due to the dissimilarities between the participants already existing prior to the experiment (degree of competences achieved in previous courses, innate testing abilities, etc.), there may exist variability between different participants applying the same treatment. Therefore, we opted for a \textit{crossover} \textbf{design}, as described by Kuehl~\cite{K:00} (a within-subjects design, where each participant applies all three techniques, but different participants apply the techniques in a different order) to prevent  dissimilarities between participants and technique application order from having an impact on results. The design of the experiment is shown in Table~\ref{tab:design}.

\begin{table}[!hbt]
  \caption{Experimental Design}
  \label{tab:design}
  \begin{tabular}{lcccccccccc}
    \toprule    
    \textbf{Program} & & \multicolumn{3}{c}{Ntree} & \multicolumn{3}{c}{Cmdline} &  \multicolumn{3}{c}{Nametbl}  \\
    \midrule
    \textbf{Period} & & \multicolumn{3}{c}{Day 1} & \multicolumn{3}{c}{Day 2} & \multicolumn{3}{c}{Day 3} \\
    \midrule
    \textbf{Technique} & \textbf{N} & CR & BT & EP & CR & BT & EP & CR & BT & EP \\
    \midrule
    \textit{Group 1} & 6 & X & - & - & - & X & - & - & - & X \\
    \textit{Group 2} & 5 & X & - & - & - & - & X & - & X & - \\
    \textit{Group 3} & 5 & - & X & - & - & - & X & X & - & - \\
    \textit{Group 4} & 5 & - & X & - & X & - & - & - & - & X \\
    \textit{Group 5} & 5 & - & - & X & X & - & - & - & X & - \\
    \textit{Group 6} & 6 & - & - & X & - & X & - & X & - & - \\
  \bottomrule
\end{tabular}
\end{table}

The \textbf{experimental procedure} takes place during seven weeks, and is summarised in Table~\ref{tab:procedure}. The first three weeks there are training sessions in which participants learn how to apply the techniques and practice with them. Training sessions take place twice a week (Tuesdays and Thursdays) and each one lasts 2 hours. Therefore, training takes 12 hours (2 hours/session x 2 sessions/week x 3 weeks). Participants are first taught the code review technique, then white-box and finally black-box. The training does not follow any particular order, but the one we have found best to meet the learning objectives of the course.

\begin{table}[!hbt]
  \caption{Experimental Procedure}
  \label{tab:procedure}
  \begin{tabular}{lccccccc}
    \toprule    
    \textbf{Type} & \multicolumn{3}{c}{Training} & Free & \multicolumn{3}{c}{Operation} \\
    \cmidrule{2-4} \cmidrule{6-8}
    \textbf{Week} & W1 & W2 & W3 & W4 & W5 & W6 & W7 \\
    \midrule
    \textbf{Activity} &  CR &  BT &  EP & Exercises & ntree & cmdline & nametbl \\
	\textbf{N. Sessions} & 2 & 2 & 2 & - & 1 & 1 & 1 \\
	\textbf{Session duration} & 2h & 2h & 2h & - & 4h & 4h & 4h \\
	\textbf{Total time} & 4h & 4h & 4h & 4h & 4h & 4h & 4h \\

  \bottomrule
\end{tabular}
\end{table}

The following week there are no lectures, and students are asked to practice with the techniques. For this purpose, they are given 3 small programs in C (that contain faults), and are asked to apply a given technique on each program (all students apply the same technique on the same training program). The performance on these exercises is used for grading purposes.

The other three are experiment execution weeks. Each experiment execution session takes place once a week (Fridays) and lasts four hours. This is equivalent to there being no time limit, as participants can complete the task in less time. Therefore, experiment execution takes 12 hours (4 hours/session x 1 session/week x 3 weeks). Training sessions take place during lecture hours and experiment execution sessions take place during laboratory hours. Those weeks in which there are lectures, there is no laboratory and vice versa. The time used for the controlled experiment is the corresponding one assigned to the course in which the study is embedded. No extra time is used. 

In each session, participants apply the techniques and, for equivalence partitioning and branch testing, run test cases too. They report application of technique, and generated test cases and failures (for the testing techniques) or faults (for the code review technique).

At the end of the last experiment execution session (after applying the last technique), participants are surveyed about their perceptions of the techniques that they applied. They must return their answer before the following Monday, to guarantee that they remember as much as possible about the tasks performed.

\subsection{Experimental Objects}
\label{sec:experimental_objects}

Program is a \textbf{blocking variable}. It is not a factor, because the goal of the experiment is not to study the programs, but the code evaluation techniques. However, it is a blocking variable, because we are aware that programs could be influencing results. The experiment has been designed to cancel out the influence of programs. Every participant applies each technique in a different program, and each technique is applied on different programs (by different participants). Additionally, the program by technique interaction is later analysed.

The experiment uses three similar \textit{programs}, written in C (used in other empirical studies about testing techniques like the ones performed by Kamsties \& Lott~\cite{KL:95} or Roper \textit{et al.}~\cite{RWM:97}):

\begin{itemize}

    \item \textit{cmdline}: parser that reads the input line and outputs a summary of its contents. It has 239 executable LOC and a cyclomatic complexity of 37.

    \item \textit{nametbl}: implementation of the data structure and operations of a symbol table. It has 230 executable LOC and a cyclomatic complexity of 27.

    \item \textit{ntree}: implementation of the data structure and operations of an n-ary tree. It has 215 executable LOC and a cyclomatic complexity of 31.

\end{itemize}

Appendix~\ref{app:metrics_programs} shows a complete listing of the metrics gathered by the PREST\footnote{Available at http://code.google.com/p/prest/} tool~\cite{KRBTC:09} on the correct programs (before faults were injected). Although the purpose of the programs is different, we can see that most of the metrics obtained by PREST are quite similar, except Halstead metrics, which are greater for ntree. At the same time, cmdline is slightly larger and more complex than the other two.

Each program has been seeded with seven faults (some, but not all, are the same faults as used in previous experiments run on these programs), and there are 2 versions of each faulty program. All faults are conceptually the same in all programs (eg., a variable initialisation is missing). Some faults occurred naturally when the programs were coded, whereas others are typical programming faults. All faults:

\begin{itemize}

	\item Cause observable failures.
	
	\item Can be detected by all techniques.
	
	\item Are chosen so that the programs fail only on some inputs. 
	
	\item No fault conceals another\footnote{One of the versions in one of the programs contains only six faults. Due to a mistake we made, one of the failures was concealed by another.}. 
	
	\item There is a one-to-one correspondence between faults and failures.

\end{itemize}

Note, however, that it is possible that a participant generates two (or more) test cases that exercise the same seeded fault, and therefore produce the same failure. Participants have been advised to report these failures (the same failure exercised by two or more different test cases) as a single one. For example, there is a fault in program ntree in the function in charge of printing the tree. This causes the failure that the tree is printed incorrectly. Every time a participant generates a test case that prints the tree (which is quite often, as this function is useful to check the contents of the tree at any time), the failure will be shown.

Some examples of the seeded faults and their corresponding failures are:

\begin{itemize}

\item Variable not initialised. The associated failure is that the number of input files is printed incorrectly in cmdline.

\item Incorrect boolean expression in a decision. The associated failure is that the program does not output error if the second node of the “are siblings” function does not belong to the tree.

\end{itemize}

\subsection{Participants}

 The 32 \textbf{participants} of the original study were fifth-(final)year undergraduate computer science students taking the elective Software Verification and Validation course at the Universidad Politécnica de Madrid. The students have gone through 2 courses on Software Engineering of 6 and 12 credits respectively. They are trained in SE, have strong programming skills, have experience programming in C, have participated in small size development projects\footnote{They have participated in development projects in teams (as in the Artificial Intelligence, Compiler and Operating Systems courses).}, and have little or no professional experience. So, they should not been considered unexperienced in programming, but good proxys of junior programmers.

They have not formal training in any code evaluation techniques (including the ones involved in the study), as this is the course in which they are taught them. Since they have had previous coding assignments, they might have done testing previously but informally. As a consequence, they might have acquired some intuitive knowledge on how to test/review programs (developing their own techniques or procedures that could resemble the techniques), but they have never learned the techniques formally. They have never been required to do peer-reviews in coding assignments, or write test cases in the projects where they have participated. They could possibly have used assertions or informal input validation, but on their own (never under request, and have not previously been taught how to do it).

All participants have a homogeneous background. The only differences could be due to the level of achievement of learning goals in previous courses, or innate ability for testing. The former could have been determined by means of scores in previous courses (which was not possible). The latter was not possible to measure. Therefore, we have not deemed necessary to do some kind of blocking, and just performed simple randomisation.

Therefore, the sample used represents developers with little or no previous experience on code evaluation techniques (novice testers). The use of our students is appropriate in this study on several grounds: 

\begin{itemize}

    \item We want to rule out any possible influence of previous experience on code evaluation techniques. Therefore, participants should not have any preconceived ideas or opinions about the techniques (including having a favourite one). 
    
    \item Falessi \textit{et al.} \cite{FJWTMJO:17} suggest that it is easier to induce a particular behaviour among students. More specifically, reinforce a high level of adherence to the treatment by experimental subjects applying the techniques.
        
    \item Students are used to make predictions during development tasks, as they are continually undergoing assessment on courses related with programming, SE, networking, etc. 
    
\end{itemize}

Having said that, since our participants are not practitioners, their opinions are not based on any previous work experience on testing, but on their experience on informally testing programs for some years (they are in 5th year of a 5-year CS bachelor). Additionally, as part of the V\&V training, our participants are asked to practice in small programs with the techniques used in the experiment.

According to Falessi \textit{et al.} \cite{FJWTMJO:17}, we (SE experiments) tend to forget practitioners’ heterogeneity. Practitioners have different academic backgrounds, SE knowledge and professional experience. For example, a developer without a computer science academic background might not have knowledge about testing techniques. We assume that for this exploratory study, the characteristics of the participants are a valid sample for developers that have little or no experience on code evaluation techniques and are junior programmers.

\subsection{Data Analysis}

The analyses conducted in response to the research questions, are explained below\footnote{All analyses are performed using IBM SPSS v26.}. Table~\ref{tab:analysis_summary} summarises the statistical tests used to answer each research question. First we report the analyses (descriptive statistics and hypothesis testing) of the controlled experiment.

\begin{table}[!hbt]
  \caption{Statistical Tests Used to Answer Research Questions}
  \label{tab:analysis_summary}
  \begin{tabular}{llll}
    \toprule

     RQ & Statistical Test & Test Description & Checks \\ \midrule

	 1.1 & $\chi^2$ goodness-of-fit & \multirow{1}{3.8cm}{How well a theoretical distribution fits a empirical distribution} & \multirow{1}{3.8cm}{All 3 techniques are equally frequently perceived as most effective} \\
	  & & & \\ 
	 & & & \\ \midrule
	 
	 1.2 &  Cohen's kappa  &  \multirow{1}{3.8cm}{Agreement for categorical variables when 2 raters classify different objects} & \multirow{1}{3.8cm}{Match between perceptions and reality} \\
 	 & & & \\
	 & & & \\
	 & Stuart-Maxwell & \multirow{1}{3.8cm}{Changes in the proportion among raters’ agreement (marginal homogeneity)} & \multirow{1}{3.8cm}{Bias in perceptions} \\
	 & & & \\
	 & & & \\
	 & McNemar-Bowker & \multirow{1}{3.8cm}{Symmetry of the associated contingency table} & \multirow{1}{3.8cm}{Bias in perceptions}\\
	 & & & \\ \midrule

	 1.3 & Krippendorff's alpha & \multirow{1}{3.8cm}{Agreement for variables when N raters classify different objects} & \multirow{1}{3.8cm}{Participants find similar amount of defects for techniques/programs} \\
	 & & & \\
	 & & & \\ \midrule

	 1.4 & 1-way ANOVA or & \multirow{1}{3.8cm}{Compare three or more groups} & \multirow{1}{3.8cm}{Different techniques show same mismatch cost/project loss}\\
	 1.5 & Kruskall-Wallis & & \\
	 &  & & \\
	           
     \bottomrule  
\end{tabular}
\end{table}

To examine \textit{participants' perceptions} (RQ1.1), we report the frequency of each technique (percentage of participants that perceive each technique as the most effective). Additionally, we determine whether all three techniques are equally frequently perceived as being the most effective. We test the null hypothesis that the frequency distribution of the perceptions is consistent with a discrete uniform distribution, i.e., all outcomes are equally likely to occur. To do this, we use a chi-square ($\chi^2$) goodness-of-fit test.

To examine if \textit{participants' perceptions predict their testing effectiveness} (RQ1.2), we use Cohen's kappa coefficient along with its 95\% confidence interval---calculated using bootstrap. Cohen's kappa coefficient (\(\kappa\)) is a statistic that measures agreement for qualitative (categorical) variables when 2 raters are classifying different objects (units). It is calculated on the corresponding contingency table generated. Table~\ref{tab:contingency_table} shows an example of a contingency table. Cells contain the frequencies associated to each pair of classes.

\begin{table}[!hbt]
  \caption{Example of Contingency Table. It is used to calculate Kappa, and perform Stuart-Maxwell's and McNemar-Bowker's tests}
  \label{tab:contingency_table}
  \begin{tabular}{ccccc}
    \toprule    
    & & \multicolumn{3}{c}{Rater 1} \\
    \cmidrule{3-5}	
    & & Class A & Class B & Class C \\
	& Class A & X$_{AA}$ & X$_{AB}$ & X$_{AC}$ \\
	Rater 2 & Class B & X$_{BA}$ & X$_{BB}$ & X$_{BC}$ \\
	& Class C & X$_{CA}$ & X$_{CB}$ & X$_{CC}$\\

  \bottomrule
\end{tabular}
\end{table}

Kappa is generally thought to be a more robust measure than simple percent agreement calculation, since it takes into account the agreement occurring by chance. It is not the only coefficient that can be used to measure agreement. There are others, like Krippendorff's alpha, which is more flexible, as can be used in situations where there are more than 2 raters, or the response variable is in an interval or ratio scale. However, in our particular situation, where there are 2 raters, data in nominal scale and no missing data, Kappa behaves similarly to Krippendorff's alpha~\cite{BCMS:99},~\cite{ZCMK:16}.

Kappa is a number from -1 to 1. Positive values are interpreted as agreement, while negative values are interpreted as disagreement. There is still some debate about how to interpret kappa. Different authors have categorised detailed ranges of values for kappa that differ with respect to the degree of agreement that they suggest (see Table~\ref{tab:kappa}). According to  scales by Altman\cite{A:91} and Landis \& Koch~\cite{LK:77}, 0.6 is the value as of which there is considered to be agreement. Fleiss et al.~\cite{FLP:03} lower this value to 0.4. Each branch of science should establish its kappa value. As there are no previous studies that specifically address the issue of which is the most appropriate agreement scale and threshold for SE, and different studies in SE have used different scales\footnote{For example, Octaviano et al. \cite{OFMF:15} use Landis \& Koch, but Massey et al. \cite{MOA:15} use Fleiss et al. as we do.}, we use Fleiss et al.'s more generous scale as our baseline.

\begin{table}[!hbt]
  \caption{Interpretation of Kappa Values. Negative values are interpreted like positive values, but meaning disagreement instead of agreement}
  \label{tab:kappa}
  \begin{tabular}{llll}
    \toprule    
    Kappa & Landis \& Koch \cite{LK:77} & Altman \cite{A:91} &  Fleiss et al. \cite{FLP:03} \\
    \midrule
    0 & Poor & Poor &  Poor \\
    \cmidrule{1-2}
    0.01--0.20 & Slight & &   \\
    \cmidrule{1-3}
    0.21--0.40 & Fair & Fair &   \\
    \midrule
    0.41--0.60 & Moderate & Moderate &  Fair to good \\
    \cmidrule{1-3}
    0.61--0.75 & Substantial & Good &   \\
    \cmidrule{1-1}  \cmidrule{4-4}
    0.76--0.80 & & & Excellent \\
    \cmidrule{1-3} 
    0.81--1.00 & Almost perfect & Very good &   \\
  \bottomrule
\end{tabular}
\end{table}

We measure the agreement between the technique perceived as most effective by a participant, and the most effective technique for that participant for all participants. Therefore, we have 2 raters (perceptions and reality), three classes (BT, EP and CR), and as many units to be classified as participants. 

Since there could be agreement for some but not all techniques, we also measure kappa for each technique separately (kappa per category), following the approach described in~\cite{E:00}. It consists of collapsing the corresponding contingency table. Table~\ref{tab:collapsed_table} shows the collapsed contingency table for Class A from Table~\ref{tab:contingency_table}. Note that a collapsed table is always a 2x2 table.

\begin{table}[!hbt]
  \caption{Example of Collapsed Contingency Table. It is used to calculate partial kappa}
  \label{tab:collapsed_table}
  \begin{tabular}{cccc}
    \toprule    
    & & \multicolumn{2}{c}{Rater 1} \\
    \cmidrule{3-4}	
    & & Class A & Other \\
	Rater 2 & Class A & X$_{AA}$ & X$_{AB}+$X$_{AC}$ \\
	& Other & X$_{BA}+$X$_{CA}$ & X$_{BB}+$X$_{BC}+$X$_{CB}+$X$_{CC}$ \\
  \bottomrule
\end{tabular}
\end{table}

In the event of disagreement, we also study the type of mismatch between perceptions and reality---whether the disagreement leads to some sort of bias in favour of any of the techniques. To do this, we use the respective contingency table to run Stuart-Maxwell's test of marginal homogeneity (testing the null hypothesis that the distribution of preferences match reality) and the McNemar-Bowker test for symmetry (testing the null hypothesis of symmetry) as explained in~\cite{E:00}. The hypothesis of marginal homogeneity corresponds to equality of row and column marginal probabilities in the corresponding contingency table. The test for symmetry determines whether observations in cells situated symmetrically about the main diagonal have the same probability of occurrence. In a 2x2 table, symmetry and marginal homogeneity are equivalent. In larger tables, symmetry implies marginal homogeneity, but the reciprocal is not true\footnote{For this reason, we need to check both.}.

Since we have injected only 7 defects in each program, there exists the possibility that if no agreement is found between perceptions and reality, it could be due to the fact that \textit{participants find a similar amount of defects for all three (or pairs of) techniques} (RQ1.3). If this is the case, then it would be difficult for them to choose the most effective technique. To check this, we will run agreement on the effectiveness obtained by participants using different techniques. Therefore we have 3 raters (techniques) and as many units as participants. This will be done with all participants, and with participants in the same experiment group, for every group; for all techniques, and for pairs of techniques. Note that kappa can no longer be used, as we are seeking agreement on interval data. For this reason, we will use Krippendorff's alpha~\cite{HK:07} along with its 95\% confidence interval---calculated using bootstrap, and the KALPHA macro for SPSS\footnote{Retrieved from: http://afhayes.com/spss-sas-and-mplus-macros-and-code.html}. 

To examine the \textit{mismatch cost} (RQ1.4) and \textit{project loss} (RQ1.5), we report the cost of the mismatch (when it is greater than zero for RQ1.4 and in all cases for RQ1.5), associated with each technique as explained in Section~\ref{sec:constructs}. To discover whether there is a relationship between the technique perceived as being the most effective and the mismatch cost and project loss, we apply a one-way ANOVA test or a medians Kruskall-Wallis test for normal and non-normal distributions, respectively along with visual analyses (scatter plots).

\section{Original Study: Validity Threats}
\label{sec:threats_it1}

Based on the checklist provided by Wohlin \textit{et al.} \cite{WRHORW:14}, the relevant threats to our study are next described.

\subsection{Conclusion Validity} 

\begin{enumerate}

\item \textit{Random heterogeneity of participants}. The use of a within-subjects experimental design ruled out the risk of the variation due to individual differences among participants being larger than the variation due to the treatment.

\end{enumerate}

\subsection{Internal Validity} 

\begin{enumerate}

\item \textit{History and maturation}:

	\begin{itemize}
		
		\item Since participants apply different techniques on different artefacts, learning effects should not be much of a concern.
		
		\item Experimental sessions take place on different days. Given the association of grades to performance in the experiment, we expect students will try to do better on the following day, causing that the technique applied the last day gets a better effectiveness. To avoid this, different participants apply techniques in different orders. This way we cancel out the threat due to order of application (avoiding that a given technique gets benefited from the maturation effect). In any case, an analysis of the chosen techniques per day is done to study maturation effect.

	\end{itemize}

\item \textit{Interactions with selection}. Different behaviours in different technique application groups are ruled out by randomly assigning participants to groups. However, we will check it analysing the behaviour of groups.

\item \textit{Hypothesis guessing}. Before filling in the questionnaire, participants in the study were informed about the goal of the study only partially. We told them that we wanted to know their preferences and opinions, but they were not aware of our research questions. In any case, if this threat is occurring, it would mean that our results for perceptions are the best possible ones, and therefore would set an upper bound.

\item \textit{Mortality}. The fact that several participants did not give consent to participate in the study has affected the balance of the experiment.

\item \textit{Order of Training}. Techniques are presented in the following order: CR, BT and EP. If this threat had taken place, then CR would be the most effective (or their favourite).

\end{enumerate}

\subsection{Construct Validity}

\begin{enumerate}

\item \textit{Inadequate preoperational explanation of cause constructs}. Cause constructs are clearly defined thanks to the extensive training received by participants on the study techniques.

\item \textit{Inadequate preoperational explanation of effect constructs}. The question being asked is totally clear and should not be subject to possible misinterpretations. However, since the perception is subjective, there exists the possibility that the question asked is interpreted differently by different participants, and hence, perceptions are related to how the question is interpreted. This issue should be further investigated in future studies.

\end{enumerate}

\subsection{External Validity}

\begin{enumerate}

\item \textit{Interaction of setting and treatment}. We tried to make the faults seeded in the programs as representative as possible of reality.

\item \textit{Generalisation to other subject types}. As we have already mentioned, the type of subjects our sample represents are developers with little or none previous experience in testing techniques and junior programmers. The extent to which the results obtained in this study can be generalised to other subject types needs to be investigated.

\end{enumerate}

Of all threats listed, the only one that could affect the validity of the results of this study in an industrial context is the one related to generalisation to other subject types.

\section{Original Study: Results}
\label{sec:results_it1}

Of the 32 students participating in the experiment, nine did not complete the questionnaire\footnote{Meaning they were not giving consent to participate in the study.} and were removed from the analysis. Table~\ref{tab:balance1} shows the balance of the experiment before and after participants submitted the questionnaire. We can see that G6 is the most affected group, with 4 missing people.

\begin{table}[!hbt]
  \caption{Balance Before and After Submitting the Questionnaire in the Original Study}
  \label{tab:balance1}
  \begin{tabular}{lcccccc}
    \toprule    
     Group & G1 & G2 & G3 & G4 & G5 & G6 \\    
    \midrule      
     Initial & 6 & 5 & 5 & 5 & 5 & 6 \\
     Final & 5 & 4 & 5 & 3 & 4 & 2 \\
    \bottomrule
\end{tabular}
\end{table}

Appendix~\ref{app:experiment_it1} shows the analysis of the experiment. The results show that program and technique are statistically significant (and therefore are influencing effectiveness), while group and the technique by program interaction are not significant.

As regards the techniques, EP shows a higher effectiveness, followed by BT and then by CR. These results are interesting, as all techniques are able to detect all defects. Additionally, more defects are found in ntree compared to cmdline and nametbl, where the same amount of defects are found. Note that ntree is the program applied the first day, has the highest Halstead metrics, and it is not the smallest program or the one with lowest complexity.

These results suggest that:

\begin{itemize}

	\item There is no maturation effect. The program where highest effectiveness is obtained is the one used the first day.

	\item There is no interaction with selections effect. Group is not significant.

	\item Mortality does not affect experimental results. The analysis technique used (Linear Mixed-Effects Models) is robust to lack of balance.
	
	\item Order of training could be affecting results. The highest effectiveness is obtained in the last technique taught, while the lowest effectiveness is obtained in the first technique taught. This suggests that techniques taught last are more effective than techniques taught first. This could be due to participants remembering better last techniques.
	
	\item Results cannot be generalised to other subject types.
	
\end{itemize}

\subsection{\label{sec:RQ11}RQ1.1: Participants’ Perceptions}

Table~\ref{tab:perci1} shows the percentage of participants that perceive each technique to be the most effective. We cannot reject the null hypothesis that the frequency distribution of the responses to the questionnaire item (\textit{Using which technique did you detect most defects?}) follows a uniform distribution\footnote{In a uniform distribution, 33.3\% of participants should choose each technique.} ($\chi^2$(2,N=23)=2.696, p=0.260). This means that the number of participants perceiving a particular technique as being more effective cannot be considered different for all three techniques. {\bf Our data do not support the conclusion that techniques are differently frequently perceived as being the most effective}.

\begin{table}[!hbt]
  \caption{Participants' Perceptions of Technique Effectiveness in the Original Study}
  \label{tab:perci1}
  \begin{tabular}{ccccc}
    \toprule    
     N & CR & BT & EP & Result \\
    \midrule      
    23 & 17.39\% & 43.48\% & 39.13\% & CR=BT=EP \\
    \bottomrule
\end{tabular}
\end{table}

\subsection{RQ1.2: Comparing Perceptions with Reality}

Table~\ref{tab:agreementi1} shows the value of kappa along with its 95\% confidence interval (CI), overall and for each technique separately. We find that all values for kappa with respect to the questionnaire item (\textit{Using which technique did you detect most defects?}) are consistent with lack of agreement ($\kappa$$<$0.4, poor). Although the upper bound of the 95\% CIs show agreement, 0 belongs to all 95\% CI, meaning that agreement by chance cannot be ruled out. Therefore, {\bf our data do not support the conclusion that participants correctly perceive the most effective technique for them}.

\begin{table}[!hbt]
  \caption{Agreement between Perceived and Real Technique Effectiveness in the Original Study (N=23)}
  \label{tab:agreementi1}
  \begin{tabular}{lccc}
    \toprule    
      & & \multicolumn{2}{c}{95\% Confidence Interval} \\      \cmidrule{3-4}
      & Kappa value & Lower bound & Upper bound \\ \midrule      
 	  Overall & 0.245 & -0.072 & 0.557 \\	
 	  CR & 0.395 & -0.131 & 0.832 \\	
 	  BT & 0.175 & -0.232 & 0.585 \\	
 	  EP & 0.225 & -0.146 & 0.566 \\	

    \bottomrule
\end{tabular}
\end{table}

It is worth noting that agreement is higher for the code review technique (the upper bound of the 95\% CI in this case shows excellent agreement). This could be attributed to participants being able to remember the actual number of defects identified in code reading whereas for testing techniques they only wrote the test cases. On the other hand, participants do not know the number of defects injected in each program.

As lack of agreement cannot be ruled out, we examine whether the perceptions are biased. The results of the Stuart-Maxwell test show that the null hypothesis of existence of marginal homogeneity cannot be rejected ($\chi^2$(2,N=23)=1.125, p=0.570). This means that we cannot conclude that perceptions and reality are differently distributed. Taking into account the results reported in Section ~\ref{sec:RQ11}, this would suggest that, in reality, techniques cannot be considered the most effective a different number of times\footnote{Note that the fact that all three techniques are classed as the most effective the same number of times is not incompatible with there being techniques that are more effective than others.}. Additionally, the results of the McNemar-Bowker test show that the null hypothesis of existence of symmetry cannot be rejected ($\chi^2$(3,N=23)=1.286, p=0.733). This means that we cannot conclude that there is directionality when participants' perceptions are wrong. These two results suggest that participants are not differently mistaken about one technique as they are about the others. {\bf Techniques are not differently subject to misperceptions}.

\subsection{RQ1.3: Comparing the Effectiveness of Techniques}

We are going to check if misperceptions could be due to participants detecting the same amount of defects with all three techniques, and therefore being impossible for them to make the right decision. Table~\ref{tab:agreementKti1} shows the value and 95\% CI of Krippendorff's $\alpha$, overall and for each pair of techniques, for all participants and for every design group (participants that applied the same technique on the same program) separately, and Table~\ref{tab:agreementKpi1} shows the value and 95\% CI of Krippendorff's $\alpha$, overall and for each program/session. For values with all participants, we can rule out agreement, as the upper bound of the 95\% CIs are consistent with lack of agreement ($\alpha$$<$0.4), except for the case of EP-BT and nametbl-ntree for which the upper bound of the 95\% CIs are consistent with fair to good agreement. However, even in this two cases, 0 belongs to the 95\% CIs, meaning that agreement by chance cannot be ruled out. This means that {\bf participants do not obtain similar effectiveness values when applying the different techniques (testing the different programs) so as to be difficult to discriminate among techniques/programs}.

\begin{table}[!hbt]
  \caption{Agreement between Percentage of Defects Found with Each Technique in the Original Study}
  \label{tab:agreementKti1}
  \begin{tabular}{lclccc}
    \toprule    
      & & & Krippendorff's & \multicolumn{2}{c}{95\% Confidence Interval} \\ \cmidrule{5-6}
      Sample & N & &  alpha value & Lower bound & Upper bound \\ \midrule      
 
 	  All & 23 & CR-BT & -0.2837 & -0.7285 & 0.1711 \\	
 	  & & CR-EP & -0.4352 & -1.0000 & 0.2524 \\	
 	  & & EP-BT & -0.1078 & -0.9040 & 0.5512 \\	
 	  & & CR-BT-EP & -0.2203 & -0.6120 & 0.0978 \\	\midrule   
 	  
 	  G1 & 5 & CR-BT & -0.1313 & -0.7848 & 0.4218 \\
 	  & & CR-EP & -0.4062 & -1.0000 & 0.5784 \\
 	  & & EP-BT & -0.1730 & -0.9193 & 0.5734 \\
 	  & & CR-BT-EP & -0.1170 & -0.7287 & 0.3989 \\ \midrule 
 
 	  G2 & 4 & CR-BT & -0.1666 & -1.0000 & 0.5461 \\
 	  & & CR-EP & -0.0722 & -0.4052 & 0.2483 \\
 	  & & EP-BT & 0.4862 & -0.0031 & 0.9755 \\
 	  & & CR-BT-EP & -0.0151 & -0.4875 & 0.4199 \\ \midrule 

	  G3 & 5 & CR-BT & -0.2956 & -0.7273 & 0.1017 \\
 	  & & CR-EP & -0.4540 & -1.0000 & 0.6506 \\
 	  & & EP-BT & -0.1368 & -1.0000 & 0.7738 \\
 	  & & CR-BT-EP & -0.2289 & -0.8099 & 0.2888 \\ \midrule 

	  G4 & 3 & CR-BT & -0.1600 & -1.0000 & 1.0000 \\
 	  & & CR-EP & -0.1600 & -1.0000 & 1.0000  \\
 	  & & EP-BT & Error & Error & Error \\
 	  & & CR-BT-EP & -0.0943 & -1.0000 & 0.8490 \\ \midrule 
 	  
	  G5 & 4 & CR-BT & -0.4789 & -1.0000 & 0.6056 \\
 	  & & CR-EP & -0.6448 & -1.0000 & -0.3768 \\
 	  & & EP-BT & -0.4260 & -1.0000 & 0.7408 \\
 	  & & CR-BT-EP & -0.3095 & -0.8496 & 0.2306 \\ \midrule 

	  G6 & 2 & CR-BT & -0.1029 & -1.0000 & 0.9559 \\
 	  & & CR-EP & -0.2931 & -1.0000 & 0.9483 \\
 	  & & EP-BT & -0.1029 & -1.0000 & 0.9559 \\
 	  & & CR-BT-EP & -0.1437 & -1.0000 & 0.9447 \\
	  
    \bottomrule
\end{tabular}
\end{table}

Furthermore, kappa values are negative, which indicates disagreement. This is good for the study, as it means that \textbf{participants should be able to discriminate among techniques, and lack of agreement cannot be attributed to a problem of being impossible to discriminate among techniques}.

As regards the results for groups, although $\alpha$ values are negative\footnote{Except in the case of Group 2 where there is agreement for the EP-BT techniques. Since this is the only agreement found we think it could be spurious.}, the 95\% CIs are too wide to show reliable results (due to small sample size). Note that in most of the cases they range from existence of disagreement in the lower bound ($\alpha$$<$-0.4) to the existence of agreement in the upper bound ($\alpha$$>$0.4).

\begin{table}[!hbt]
  \caption{Agreement between Percentage of Defects Found with Each Program in the Original Study (N=23)}
  \label{tab:agreementKpi1}
  \begin{tabular}{lccc}
    \toprule    
      & Krippendorff's & \multicolumn{2}{c}{95\% Confidence Interval} \\ \cmidrule{3-4}
      &  alpha value & Lower bound & Upper bound \\ \midrule      
 	  cmdline-nametbl & -0.3301 & -1.0000 & 0.3137 \\	
 	  cmdline-ntree & -0.1801 & -0.7331 & 0.2787 \\	
 	  nametbl-ntree & -0.1808 & -0.9570 & 0.4400 \\	
 	  cmdline-nametbl-ntree & -0.2203 & -0.5933 & 0.1300 \\	
    \bottomrule
\end{tabular}
\end{table}

\subsection{RQ1.4: Cost of Mismatch}

\begin{figure}[!hbt]
\includegraphics[width=3.5in]{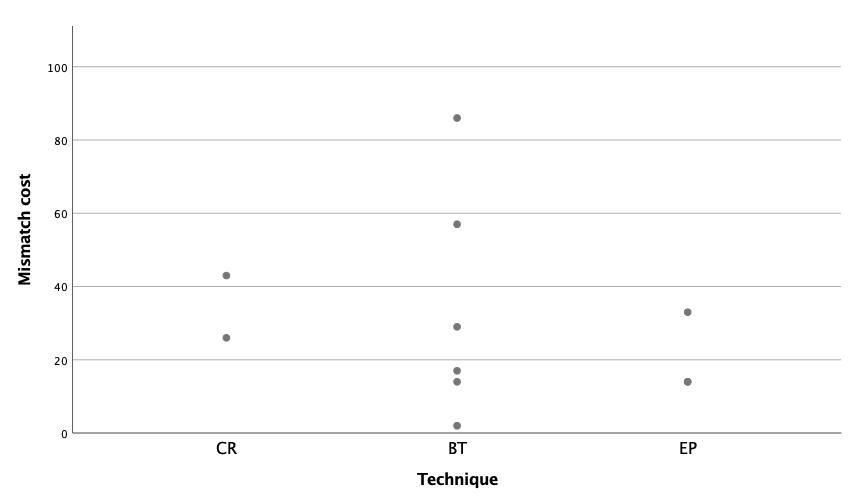}
\caption{Scatterplot for observed mismatch cost in the original study. Datapoints correspond to the mismatch cost in Table~\ref{tab:s1_cost}}
\label{fig:spmcit1}
\end{figure}

Table~\ref{tab:s1_cost} and Figure~\ref{fig:spmcit1} show the cost of mismatch. We can see that the EP technique has fewer mismatches compared to the other two. Additionally, the mean and median mismatch cost is smaller. On the other hand, the BT technique has more mismatches, and a higher dispersion. The results of the Kruskal-Wallis test reveal that we cannot reject the null hypothesis of techniques having the same mismatch cost (H(2)=0.685, p=0.710). This means that we cannot claim a difference in mismatch cost between the techniques. The estimated mean mismatch cost is 31pp (median 26pp).

\begin{table}[!hbt]
  \caption{Observed Reduction in Technique Effectiveness for Mismatch. Column 2 shows the number of mismatches out of the total number of participants who perceived the technique as being most effective. Column 3 shows the cost for each mismatch. Columns 4-6 shows the mean and median (in percentage points), and standard deviation for mismatch cost}
  \label{tab:s1_cost}
  \begin{tabular}{cclccc}
    \toprule
     & & & \multicolumn{3}{c}{Cost} \\
     \cmidrule{4-6}  
     Technique & No. Mismatches & Mismatch Cost & Mean & Median & Std. Deviation \\
    \midrule     
     CR & 2(4) & 26pp; 43pp & 35pp & 35pp & 12 \\
     BT & 6(10) & 2pp; 14pp; 17pp; & 34pp & 23pp & 31 \\
     & & 29pp; 57pp; 86pp & & & \\
     EP & 3(9) & 14pp; 14pp; 33pp & 21pp & 14pp & 11 \\
     \midrule
     TOTAL & 11(23) &  & 31pp & 26pp & 24 \\
    \bottomrule
\end{tabular}
\end{table}

These results suggest that {\bf the mismatch cost is not negligible} (31pp), {\bf and is not related to the technique perceived as most effective.} However, note that the existence of very high mismatches and few datapoints could be affecting these results.

\subsection{RQ1.5: Expected Loss of Effectiveness}

Table~\ref{tab:s1_cost_all} shows the average loss of effectiveness that should be expected in a project, where typically different testers participate, and therefore, there would be both matches and mismatches\footnote{Note that the mismatch cost is 0 when there is a match.}. Again, the results of the Kruskal-Wallis test reveal that we cannot reject the null hypothesis of techniques having the same expected reduction in technique effectiveness for a project (H(2)=1.510, p=0.470). This means we cannot claim a difference in project effectiveness loss between techniques. The mean expected loss in effectiveness in the project is estimated as 15pp\footnote{Note that the median here is not very informative. In this particular case it is 0pp. This happens when there are more matches than mismatches.}.

\begin{table}[!hbt]
  \caption{Observed Reduction in Technique Effectiveness when Considering Matches and Mismatches. Column 2 shows the number of datapoints. Column 3 shows the cost for each (mis)match. Columns 4-6 show the mean and median (in percentage points), and std. deviation for the reduction in technique effectiveness}
  \label{tab:s1_cost_all}
  \begin{tabular}{cclccc}
    \toprule
     & & & \multicolumn{3}{c}{Cost} \\
     \cmidrule{4-6}  
     Technique & N & (Mis)match Cost & Mean & Median & Std. Deviation \\
    \midrule     
     CR & 4 & 0pp; 0pp; 26pp; 43pp & 17pp & 13pp & 21 \\
     BT & 10 & 0pp; 0pp; 0pp; 0pp; 2pp; & 21pp &  8pp & 29 \\
     & & 14pp; 17pp; 29pp; 57pp; 86pp & & & \\
     EP & 9 & 0pp; 0pp; 0pp; 0pp; 0pp; & 7pp  & 0pp & 12 \\
     & & 0pp; 14pp; 14pp; 33pp & & \\
     \midrule
     TOTAL & 23 & & 15pp & 0pp & 22 \\
    \bottomrule
\end{tabular}
\end{table}

These results suggest that {\bf the expected loss in effectiveness in a project is not negligible} (15pp), {\bf and is not related to the technique perceived as most effective.} However, we must note again that the existence of very high mismatches for BT and few datapoints could be affecting these results.

\subsection{Findings of the Original Study}

 Our {\bf findings} are:
 
 \begin{itemize}
 
 	\item Participants should not base their decisions on their own perceptions, as their perceptions are not reliable and have an associated cost.
 	
 	\item We have not been able to find a bias towards one or more particular techniques that might explain the misperceptions.
 	
 	\item Participants should have been able to identify the different effectiveness of techniques.
 	
 	\item Misperceptions cannot be put down to experience. The possible drivers of these misperceptions require further research.
 
 \end{itemize}

Note that these findings cannot be generalised to other types of developers rather than those with the same profile as the ones used in this study.

\section{Replicated Study: Research Questions and Methodology}
\label{sec:methodology_it2}

We decide to further investigate the results of the original study in search of possible drivers behind misperceptions. Psychology considers that people's perceptions can be affected by personal characteristics as attitudes, personal interests and expectations. Therefore, we decide to examine participants' opinions by conducting a differentiated replication of the original study~\cite{SCVJ:08} that extends its goal as follows: 

\begin{enumerate}

\item The survey of effectiveness perception is extended to include questions on programs.

\item We want to find out whether participants' perceptions might be conditioned by their opinions. More precisely: their preferences (favourite technique), their performance (the technique that they think they applied best) and technique or program complexity (the technique that they think is easiest to apply, or the simplest program to be tested).
  
\end{enumerate}

Therefore, the replicated study reexamines RQ1 stated in the original study (this time the survey taken by participants also includes questions regarding programs), and addresses the new following research questions:

\begin{itemize}

\item \textit{RQ1.6: Are participants perceptions related to the number of defects reported by participants?}

	We want to assess if participants perceive as the most effective technique the one with which they have reported more defects.

\item \textit{RQ2: Can participants' opinions be used as predictors for testing effectiveness?}

    \begin{itemize}

    \item \textit{RQ2.1: What are participants' opinions about techniques and programs?} 
    
    We want to know if participants have different opinions about techniques or programs.

    \item \textit{RQ2.2: Do participants' opinions predict their effectiveness?} 
    
    We want to assess if the opinions that participants have about techniques (or programs) predict which one is the most effective for them.
    
    \end{itemize}

\item \textit{RQ3: Is there a relationship between participants' perceptions and opinions?} 

\begin{itemize}

    \item \textit{RQ3.1: Is there a relationship between participants' perceptions and opinions?}

   We want to assess if the opinions that participants have about techniques (or programs) are related to their perceptions. 

    \item \textit{RQ3.2: Is there a relationship between participants' opinions?}
    
      We want to assess if a certain opinion that participants have about techniques are related to other opinions. 
    
\end{itemize}

\end{itemize}

To answer these questions, we replicate the original study with students of the same course in the following academic year. This time we have 46 students. The changes made to the replication of the experiment are as follows:

\begin{itemize}

\item The questionnaire to be completed by participants at the end of the experiment is extended to include new questions. 
 The information we want to capture with the opinion questions is:

\begin{itemize}

	\item \textit{Participants performance on techniques}. With this question we are referring to process conformance. Best applied technique is the technique each participant thinks (s)he applied more thoroughly. It corresponds to OT1: Which technique did you apply best?
	
	\item \textit{Participants preferences}. We want to know the favourite technique of each participant. They one (s)he felt more comfortable with when applied. It corresponds to OT2: Which technique do you like best?

	\item \textit{Technique complexity}. We want to know the technique each participant thinks was easiest to get process conformance. It corresponds to OT3: Which technique is the easiest to apply?

	\item \textit{Program testability}. We want to know the program it was easier to test. This is, the program in which process conformance could be obtained more easily. It corresponds to OP1: Which is the simplest program?
	
\end{itemize}

Table~\ref{tab:perceps2} summarizes the survey questions. We have chosen these questions because we need to ask simple questions, that can be easily understood by participants, being at the same time meaningful. We do not want to overwhelm participants with complex questions that have lots of explanations. A complex questionnaire might discourage students to submit it.

\begin{table}[!hbt]
  \caption{Questions of Replicated Study Questionnaire}
  \label{tab:perceps2}
  \begin{tabular}{lll}
    \toprule    
    ID & Question & Aspect\\
    \midrule
    PT1 & Using which technique did you detect most defects? & Perceptions \\   
    PP1 & In which program did you detect most defects? &  \\
    \midrule
    OT1 & Which technique did you apply best? & \\
    OT2 & Which technique do you like best? &  Opinions \\
    OT3 & Which technique is the easiest to apply? & \\
    OP1 & Which is the simplest program? & \\
   \bottomrule
\end{tabular}
\end{table}

\item The program faults are changed. The original study is designed so that all techniques are effective at finding all defects injected. We choose faults detectable by all techniques so the techniques could be compared fairly. The replicated study is designed to cover the situation in which some faults cannot be detected by all techniques. Therefore, we inject some faults that techniques are not effective at detecting.
 
For example, BT cannot detect a non-implemented feature (as participants are required to generate test cases from the source code only). Likewise, EP cannot find a fault whose detection depends on the combination of two invalid equivalence classes. Therefore, in the replicated study, we inject some faults that can be detected by BT but not by EP and some faults that can be detected by EP but not by BT into each program (each program is seeded with six faults). Note that  the design is balanced: we inject the same number of faults that BT can detect, but not EP, that the opposite –EP can detect, but not BT). This change is expected to affect the effectiveness of EP and BT, which might be lower than in the original study. It should not affect the effectiveness of CR.

\item We change the program application order to further study maturation issues. The order is now: cmdline, ntree, nametbl. This change should not affect the results.

\item Participants run their own test cases. It could be that the misperceptions obtained in the original study are due to the fact that participants are not running their own test cases.

\item There are not two versions anymore but one. Faults and failures are not the goal of this study. This helps to simplify the experiment.

\end{itemize}

Table~\ref{tab:changes} shows a summary of the changes made to the study. 
 \begin{table}[!hbt]
  \caption{Changes Made to the Original Study}
  \label{tab:changes}
  \begin{tabular}{ll}
    \toprule    
     Change & Purpose \\
    \midrule
    New questions & Extend scope of study \\
    Program faults & Cover all possible scenarios \\
    Program application order & Further study maturation issues \\
    Participants run their own test cases & Improve perceptions \\
    Just one program version & Simplify experiment \\
    \bottomrule
\end{tabular}
\end{table}

 To measure technique effectiveness we proceed in the same way as in the original study. We do not rely on the reported failures, as participants could:

\begin{enumerate}

	\item Report false positives (non-real failures).

	\item Report the same failure more than once (although they were asked not to do so).
	
	\item Miss failures corresponding to faults that have been exercised by the technique, but for some reason have not been seen.
 
\end{enumerate}

We measure the new response variable (reported defects) by counting the number of faults/failures reported by each participant.

We analyse RQ2.1 in the same manner as RQ1.1, and RQ1.6, RQ2.2, RQ3.1 and RQ3.2 like RQ1.2. Table~\ref{tab:analysis_summary_it2} summarises the statistical tests used to answer each research question.

\begin{table}[!hbt]
  \caption{Statistical Tests Used to Answer New Research Questions of the Replicated Study}
  \label{tab:analysis_summary_it2}
  \begin{tabular}{llll}
    \toprule     

     RQ & Statistical Test & Test Description & Checks \\ \midrule

	 2.1 & $\chi^2$ goodness-of-fit & \multirow{1}{3.8cm}{How well a theoretical distribution fits a empirical distribution} & \multirow{1}{3.8cm}{All 3 techniques/programs are equally frequently perceived/opined as most effective} \\
	 & & & \\ 
	 & & & \\
	 & & & \\ \midrule
	 1.6 &  Cohen's kappa  &  \multirow{1}{3.8cm}{Agreement for categorical variables when 2 raters classify different objects} & \multirow{1}{3.8cm}{Match between perceptions or opinions and reality and among them} \\
 	 2.2 & & & \\
	 3.1 & & & \\
	 3.2 & Stuart-Maxwell & \multirow{1}{3.8cm}{Changes in the proportion among raters’ agreement (marginal homogeneity)} & \multirow{1}{3.8cm}{Bias in perceptions/opinions} \\
	 & & & \\
	 & & & \\
	 & McNemar-Bowker & \multirow{1}{3.8cm}{Symmetry of the associated contingency table} & \multirow{1}{3.8cm}{Bias in perceptions/opinions}\\
	 & & & \\
          
     \bottomrule  
\end{tabular}
\end{table}

\section{Replicated Study: Validity Threats}
\label{sec:threats_it2}

The threats to validity listed in the original study apply to this replicated study. Additionally, we have identified the following ones:

\subsection{Conclusion Validity} 

\begin{enumerate}

\item \textit{Reliability of treatment implementation}. The replicated experiment is run by the same researchers that performed the original experiment. This assures that the two groups of participants do not implement the treatments differently.

\end{enumerate}

\subsection{Internal Validity}

\begin{enumerate}

\item \textit{Evaluation Apprehension}. The use of students and associating their performance in the experiment with their grade in the course might explain that participants consider that their performance and not the weaknesses of the techniques explain the effectiveness of a technique.

\end{enumerate}

\subsection{Construct Validity}

\begin{enumerate}

\item \textit{Inadequate preoperational explanation of effect constructs}. Since opinions are hard constructs to operationalize, there exists the possibility that the questions appearing in the questionnaire are not interpreted by participants the way we intended to.

\end{enumerate}

\subsection{External Validity}

\begin{enumerate}

\item \textit{Reproducibility of results}. It is not clear to what extent the results obtained here are reproducible. Therefore, more replications of the study are needed. The steps that should be followed are:

	\begin{enumerate}
	
	\item Replicate the study capturing the reasons for the answers given by participants.

	\item Perform the study with practitioners with the same characteristics as the students used in this study (people with little or no experience in software testing).

	\item Explore and define what types of experience could be influencing the results (academic, professional, programming, testing, etc.).

	\item Run new studies taking into consideration increasing levels of experience.  	
	
	\end{enumerate}

\end{enumerate}

Again, of all threats affecting the replicated study, the only one that could affect the validity of the results of this study in an industrial context is the one related to generalisation to other subject types.

\section{Replicated Study: Results}
\label{sec:results_it2}

Of the 46 students participating in the experiment, seven did not complete the questionnaire\footnote{Meaning they were not giving consent to participate in the study.} and were removed from the analysis. Table~\ref{tab:balance2} shows the changes in the experimental groups due to students not participating in the study. Balance is not seriously affected by mortality---although it would have been desirable that Group 5 had at least one more participant.

\begin{table}[!hbt]
  \caption{Balance Before and After Dropouts in the Replicated Study}
  \label{tab:balance2}
  \begin{tabular}{lcccccc}
    \toprule    
     Group & G1 & G2 & G3 & G4 & G5 & G6 \\    
    \midrule      
     Initial & 8 & 8 & 8 & 8 & 7 & 7 \\
     Final & 6 & 7 & 7 & 8 & 5 & 6 \\
    \bottomrule
\end{tabular}
\end{table}

Additionally, another four participants did not answer all the questions and were removed from the analysis of the respective questions.

\subsection{RQ1: Participants’ Perceptions as Predictors}

\subsubsection{RQ1.1-RQ1.5: Comparison with Original Study Results}

Appendix~\ref{app:experiment_it2} shows the analysis of the experiment. Program is the only statistically significant variable (group, program and the program by technique interaction are not significant). In this replication, fewer defects are found in cmdline compared to nametbl and ntree, where the same amount of defects are found. Some results are in line with those obtained in the original study:

\begin{itemize}

	\item There is no interaction with selections effect. Group is not significant.

	\item Mortality does not affect experimental results. The analysis technique used (Linear Mixed-Effects Models) is robust to lack of balance.
	
	\item Results cannot be generalized to other subject types.
	
\end{itemize}

But others contradict those obtained in the original study, and therefore need further investigation:

\begin{itemize}

	\item Maturation effect cannot be ruled out. The program where lowest effectiveness is obtained is the one used the first day. 

	\item Order of training does not seem to be affecting results. All techniques show the same effectiveness. 		
	
\end{itemize}

Table~\ref{tab:perci2} shows the results of participants’ perceptions for techniques. The results are the same as in the original study ($\chi^2$(2,N=37)=3.622, p=0.164). {\bf Our data do not support the conclusion that techniques are differently frequently perceived as being the most effective.}.

\begin{table}[!hbt]
  \caption{Participants' Perceptions for Technique Effectiveness in the Replicated Study}
  \label{tab:perci2}
  \begin{tabular}{lccccc}
    \toprule     
     Question & N & CR & BT & EP &  Result \\
     \midrule      
     PT1 & 37 & 37.84\% & 18.92\% & 43.24\% &  CR=BT=EP \\
    \midrule  
\end{tabular}
\end{table}

\textbf{Our data do not support the conclusion that participants correctly perceive the most effective technique for them}. The overall and per technique kappa values and 95\% CI reported in Table~\ref{tab:agreementti2} are in line with those in the original study. This suggests that the hypothesis we elaborated in the original experiment would not be correct. For some reason, perceptions are more accurate with the CR technique.

\begin{table}[!hbt]
  \caption{Agreement between Technique Effectiveness Perceptions and Reality in Replicated Study (PT1, N=37)}
  \label{tab:agreementti2}
  \begin{tabular}{lccc}
    \toprule    
      & & \multicolumn{2}{c}{95\% Confidence Interval} \\      \cmidrule{3-4}
      & Kappa value & Lower bound & Upper bound \\ \midrule      
 	  Overall & 0.193 & -0.023 & 0.430 \\	
 	  CR & 0.364 & 0.078 & 0.628 \\	
 	  BT & -0.025 & -0.253 & 0.311 \\	
 	  EP & 0.142 & -0.176 & 0.446 \\	
    \bottomrule
\end{tabular}
\end{table}

Again as in the original study, \textbf{we have not been able to observe bias in perceptions} (Stuart-Maxwell outputs ($\chi^2$(2, N=37)=3.103,p=0.212), and McNemar-Bowker outputs ($\chi^2$(3,N=37)=3.143, p=0.370)).

Table~\ref{tab:agreementKti2} shows the value of Krippendorff's $\alpha$ and 95\% CI, overall and for each pair of techniques, for all participants and for every design group (participants that applied the same technique on the same program) separately, and Table~\ref{tab:agreementKpi2} shows the value of Krippendorff's $\alpha$ and 95\% CI, overall and for each program/session. Again, the results obtained are the same as in the original study. {\bf Participants do not obtain similar effectiveness values when applying the different techniques (testing the different programs) so as to be difficult to discriminate among techniques/programs}. 

\begin{table}[!hbt]
  \caption{Agreement between Techniques for Percentage of Defects Found in the Replicated Study}
  \label{tab:agreementKti2}
  \begin{tabular}{lclccc}
    \toprule    
      & & & Krippendorff's & \multicolumn{2}{c}{95\% Confidence Interval} \\ \cmidrule{5-6}
      Sample & N & &  alpha value & Lower bound & Upper bound \\ \midrule      

 	  All & 39 & CR-BT & 0.0462 & -0.2975 & 0.3384  \\	
 	  & & CR-EP & -0.0217 & -0.4304 & 0.3444 \\	
      &  & EP-BT & -0.0074 & -0.4577 & 0.3363 \\	
 	  & & CR-BT-EP & 0.0083 & -0.1988 & 0.2029 \\	 
 	  \midrule   
 	  
 	  G1 & 6 & CR-BT & 0.0782 & -0.9665 & 0.8771 \\
 	  & & CR-EP & -0.0447 & -1.0000 & 0.6313 \\
 	  &  & EP-BT & 0.0833 & -0.7188 & 0.6563 \\
  	  & & CR-BT-EP & 0.0368 & -0.5652 & 0.5184 \\ 
 	  \midrule 
 
  	  G2 & 7 & CR-BT & 0.0000 & -0.5000 & 0.5000 \\
 	  & & CR-EP & 0.2593 & -0.1493 & 0.6680 \\
	  & & EP-BT & 0.2642 & -0.5698 &  0.9019 \\
 	  & & CR-BT-EP & 0.1499 & -0.1969 & 0.4855 \\ 
 	  \midrule 

 	  G3 & 7 & CR-BT & 0.2500 & -0.1875 & 0.6875 \\
 	  & & CR-EP & 0.1727 & -0.4182 & 0.7045 \\
	  & & EP-BT & 0.1096 & -0.6918 & 0.6884 \\
 	  & & CR-BT-EP & 0.1958 & -0.1713 & 0.5105 \\ 
 	  \midrule 

 	  G4 & 8 & CR-BT & -0.2069 & -0.8966 & 0.3103 \\
 	  & & CR-EP & -0.1290 & -1.0000 & 0.6237 \\
	  & & EP-BT & 0.1916 & -0.4371 & 0.6407 \\
 	  & & CR-BT-EP & -0.0758 & -0.5395 & 0.3137 \\ 
 	  \midrule 
 	  
 	  G5 & 5 & CR-BT & -0.3125 & -1.0000 & 0.8125 \\
 	  & & CR-EP & -0.5672 & -1.0000  & 0.5522 \\
	  & & EP-BT & -0.6211 & -1.0000 & 0.2733 \\
 	  & & CR-BT-EP & -0.3515 & -0.9926 & 0.2723 \\ 
 	  \midrule 

 	  G6 & 6 & CR-BT & -0.3750 & -1.0000 & 0.7250 \\
 	  & & CR-EP & -0.2222 & -0.8333 & 0.3889 \\
	  & & EP-BT & 0.0833 & -0.3750 & 0.5417 \\
 	  & & CR-BT-EP & -0.1168 & -0.7993 &  0.4416  \\
	  
    \bottomrule
\end{tabular}

\end{table}

\begin{table}[!hbt]
  \caption{Agreement between Programs for Percentage of Defects Found in the Replicatd Study (N=39)}
  \label{tab:agreementKpi2}
  \begin{tabular}{lccc}
    \toprule    
      & Krippendorff's & \multicolumn{2}{c}{95\% Confidence Interval} \\ \cmidrule{3-4}
      &  alpha value & Lower bound & Upper bound \\ \midrule      
 	  cmdline-nametbl & -0.0240 & -0.4017 & 0.3040 \\	
 	  cmdline-ntree & -0.0398 & -0.4049 & 0.3095 \\	
 	  nametbl-ntree & 0.0304 & -0.3716 & 0.3851 \\	
 	  cmdline-nametbl-ntree & 0.0083 & -0.2082 & 0.2092 \\	
    \bottomrule
\end{tabular}
\end{table}

\begin{figure}[!hbt]
\includegraphics[width=3.5in]{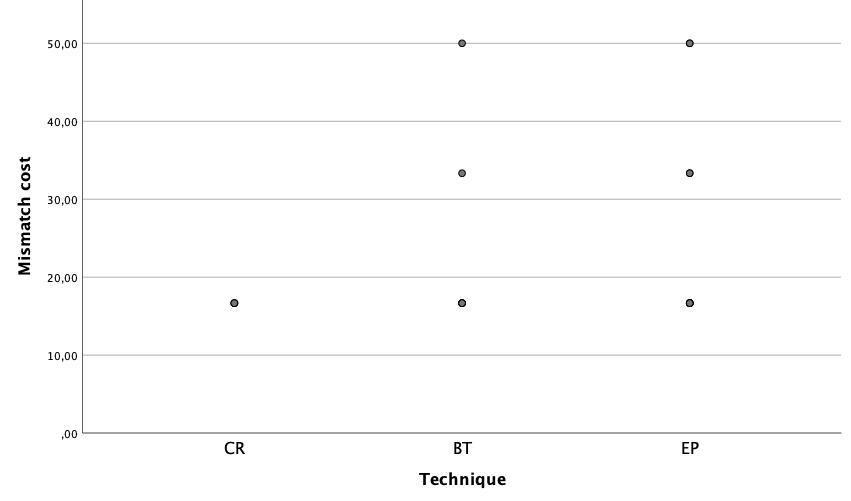}
\caption{Scatterplot for observed mismatch cost in the replicated study}.
\label{fig:spmcit2}
\end{figure}

Table~\ref{tab:s2_tcost1} and Figure~\ref{fig:spmcit2} show the cost of mismatch. As in the original study, the {\bf mismatch cost is not related to the technique perceived as being the most effective,} (Kruskal-Wallis (H(2)=2.979,p=0.226)). Also, there are about the same proportion of mismatches as in the original study (48\% of mismatches in the original study versus 51\% in the replicated study.

\begin{table}[!hbt]
  \caption{Observed Reduction in Technique Effectiveness for Mismatch. Column 2 shows the number of mismatches out of the total number of participants who perceived the technique as being most effective. Column 3 shows the cost for each mismatch. Columns 4-6 shows the mean and median (in percentage points), and standard deviation for mismatch cost}
  \label{tab:s2_tcost1}
  \begin{tabular}{cclccc}
    \toprule
     & &  & \multicolumn{3}{c}{ Cost}\\  \cmidrule{4-6}  
     Technique & No. Mismatches & Mismatch Cost & Mean & Median & Std. Deviation \\
    \midrule     
     CR & 3(14) & 17pp; 17pp; 17pp & 17pp & 17pp & 0\\
     BT & 6(7) & 17pp; 50pp; 17pp;    & 25pp &  17pp & 14\\
     & & 17pp; 17pp; 33pp & & & \\
     EP & 10(16) & 17pp; 50pp; 17pp;  & 27pp & 17pp & 14\\
     & & 17pp; 17pp; 33pp; & & & \\
     & & 17pp; 50pp; 33pp;  & & & \\
     & & 17pp & & & \\
    \midrule     
	 TOTAL & 19(37) & & 25pp & 17pp & 13\\

    \bottomrule
\end{tabular}
\end{table}

However, there are some differences with respect to the original study:

\begin{itemize}

	\item While CR had the greatest number of mismatches in the original study, now it has the smallest. The number of mismatches for BT and EP has increased with respect to the original study. 
	
	\item In the replicated study, the mismatch cost is slightly lower (25pp compared with 31pp in the original study). The mismatch cost is smaller when CR is involved. 
	
\end{itemize}

This could be due to the change in the seeded faults or just to natural variation. It should be further checked. However, it is a fact that the effectiveness of EP and BT has decreased in the replicated study, while CR has a similar effectiveness as in the original study. This suggests that {\bf the mismatch cost could be related to the faults that the program contains}. However, this issue needs to be further investigated, as we have few data points. Note that, as in the original study, the existence of few datapoints could be affecting these results.

Table \ref{tab:s2_tcost2} shows the average loss of effectiveness that should be expected in a project due to mismatch. The expected loss in effectiveness in a project is similar to the one observed in the original study (13pp), but this time it is related to the technique perceived as most effective (Kruskal-Wallis (H(2)=9.691, p=0.008)). This means that some mismatches are more costly than others. The misperception of CR as being the most effective technique has a lower associated cost (4pp) than for BT or EP (18pp). This suggests that participants’ who think CR is the most effective might be allowed to apply this technique, as, even if they are wrong, the loss of effectiveness would be negligible. However, {\bf participants should not rely on their perceptions} even in this case, since {\bf fault type could have an impact} on this result and they will never know what faults there are in the program beforehand. Note that again the existence of few datapoints could be affecting these results. Therefore, this issue needs to be further researched.

\begin{table}[!hbt]
  \caption{Observed Reduction in Technique Effectiveness when Considering Matches and Mismatches. Column 2 shows the number of datapoints. Column 3 shows the cost for each (mis)match. Columns 4-6 show the mean and median (in percentage points), and std. deviation for reduction in effectiveness in the project}
  \label{tab:s2_tcost2}
  \begin{tabular}{cclccc}
    \toprule
     &  & & \multicolumn{3}{c}{ Cost}\\  \cmidrule{4-6}  
     Technique & N & (Mis)match Cost & Mean & Median & Std. Deviation \\
    \midrule     
     CR & 14 & 0pp; 0pp; 0pp; 0pp; 0pp; 0pp; & 4pp & 0pp & 7 \\
     & & 0pp; 0pp; 0pp; 0pp; 0pp; 17pp; & & &\\
     
     & & 17pp; 17pp & & & \\
     BT & 7 & 0pp; 17pp; 50pp; 17pp; 17pp; & 21pp &  17pp & 16\\
     & & 17pp; 33pp & & & \\
     EP & 16 & 0pp; 0pp; 0pp; 0pp; 0pp; 0pp;  & 17pp  & 17pp & 17 \\
     & & 17pp; 17pp; 17pp; 17pp; 17pp;  & & \\
     & & 17pp; 33pp; 33pp; 50pp; 50pp & & \\
     \midrule     
     TOTAL & 37 &  & 13pp & 17pp & 15 \\
    \bottomrule
\end{tabular}
\end{table}

The {\bf findings of the replicated study} are:

\begin{itemize}

\item They confirm the results of the original study. 

\item A possible relationship between fault type and mismatch cost should be further investigated.

\end{itemize}
 
Since the results of both studies are similar, we have pooled the data and performed joint analyses for all research questions to overcome the problem of lack of power due to sample size. They are reported in Appendix~\ref{app:joint_analyses}. The results confirm those obtained by each study individually. This allows us to gain confidence in the results obtained.

\subsubsection{RQ1.6: Perceptions and Number of Defects Reported}

One of the conclusions of the original study was that perceived technique effectiveness could match the technique with the highest number of defects reported. Table~\ref{tab:agreement_defects} shows the value of kappa and its 95\% CI, overall and for each technique separately. We find that all values for kappa with respect to the \textit{perceived most effective technique and technique with greater number of defects reported} are consistent with lack of agreement ($\kappa$$<$0.4, poor). However, the upper bound of all 95\% CIs show agreement, and the lower bound of all 95\% CIs but BT are greater than zero. This means that {\bf although our data do not support the conclusion that participants correctly perceive the most effective technique for them, it should not be ruled out}.

 This means that {\bf participants perceptions about technique effectiveness could be related to reporting a greater number of defects with that technique}. 

\begin{table}[!hbt]
  \caption{Agreement between Perceived Most Effective Technique and Technique with Greater Number of Defects Reported in the Replicated Study (N=37)}
  \label{tab:agreement_defects}
  \begin{tabular}{lccc}
    \toprule    
      & & \multicolumn{2}{c}{95\% Confidence Interval} \\      \cmidrule{3-4}
      & Kappa value & Lower bound & Upper bound \\ \midrule      
 	  Overall & 0.289 & 0.059 & 0.518 \\	
 	  CR & 0.347 & 0.051 & 0.624 \\	
 	  BT & 0.081 & -0.229 & 0.416 \\	
 	  EP & 0.371 & 0.065 & 0.645 \\	
    \bottomrule
\end{tabular}
\end{table}

As lack of agreement cannot be ruled out, we examine whether the perceptions are biased. The results of the Stuart-Maxwell test show that the null hypothesis of existence of marginal homogeneity cannot be rejected ($\chi^2$(2,N=37)=2.458, p=0.293). This means that we cannot conclude that perceptions and reported defects are differently distributed. Additionally, the results of the McNemar-Bowker test show that the null hypothesis of existence of symmetry cannot be rejected ($\chi^2$(3,N=37)=2.867, p=0.413). This means that we cannot conclude that there is directionality when participants' perceptions do not match the technique with highest defects reported. 

The lack of a clear agreement could be due to the fact that participants do not remember exactly the number of defects found with each technique.

\subsubsection{RQ1.1-RQ1.2: Program Perceptions}

Table~\ref{tab:percpi2} shows the results of participants’ perceptions for the \textit{program in which the participants detected most defects}. We found that the same phenomenon applies to programs as to techniques. All three programs cannot be considered differently frequently perceived as being the ones where most defects were found, as we cannot reject the null hypothesis that the frequency distribution of the responses follows a uniform distribution ($\chi^2$(2,N=37)=2.649, p=0.266). {\bf Our data do not support the conclusion that programs are differently frequently perceived as having a higher percentage of defects found than the others}. This contrasts with the fact that cmdline has a slightly higher complexity and number of LOC, and that ntree shows highest Halstead metrics. We expected cmdline and/or ntree should be perceived less frequently as having a higher detection rate.

\begin{table}[!hbt]
  \caption{Participants' Perceptions of Program Effectiveness in the Replicated Study (PP1)}
  \label{tab:percpi2}
  \begin{tabular}{lccccc}
    \toprule     
     Question & N & cmdline & nametbl & ntree & Result \\
     \midrule      
      PP1 & 37 & 43.24\% & 35.14\% & 21.62\% &  Cm=Na=Nt \\     
    \bottomrule
\end{tabular}
\end{table}

However, the values for kappa in Table~\ref{tab:agreementpi2} show that there seems to be agreement overall and for cmdline and ntree ($\kappa>$0.4, fair to good and agreement by chance can be ruled out, since 0 does not belong to the 95\% CI), but not so for the nametbl program ($\kappa$=0.292, poor and agreement by chance cannot be ruled out, as 0 belongs to the 95\% CI). This means that {\bf participants do tend to correctly perceive the program in which they detected most defects}. This is striking, as it contrasts with the disagreement for techniques. Pending the analysis of the mismatch cost, it suggests that participants’ perceptions on the percentage of defects found may be reliable. This is interesting, as cmdline has a higher complexity.

\begin{table}[!hbt]
  \caption{Agreement between Program Effectiveness Perceptions and Reality in the Replicated Study (PP1, N=37)}
  \label{tab:agreementpi2}
  \begin{tabular}{lccc}
    \toprule    
      & & \multicolumn{2}{c}{95\% Confidence Interval} \\      \cmidrule{3-4}
      & Kappa value & Lower bound & Upper bound \\ \midrule      
 	  Overall &  \cellcolor{lightgray} 0.401 & 0.186 & 0.621 \\	
 	  cmdline & \cellcolor{lightgray} 0.469 & 0.208 & 0.711 \\	
 	  nametbl & 0.292 & -0.011 & 0.617 \\	
 	  ntree & \cellcolor{lightgray} 0.460 & 0.137 & 0.769 \\	
    \bottomrule
\end{tabular}

\end{table}

Since there is agreement, we are not going to study mismatch cost.

\textbf{Misperceptions do not seem to affect participants’ perception of how well they have tested a program.}

\subsection{RQ2: Participants’ Opinions as Predictors}

\subsubsection{RQ2.1: Participants’ Opinions}

Table~\ref{tab:opit2} shows the results for participants’ opinions with respect to techniques. 

\begin{table}[!hbt]
  \caption{Participants' Opinions for Techniques in the Replicated Study}
  \label{tab:opit2}
  \begin{tabular}{lccccc}
    \toprule    
     Question & N & CR & BT & EP  & Result \\
    \midrule      
     OT1 & 38 & 26.32\% & 15.79\% & 57.89\% & EP$>$(BT=CR) \\
     OT2 & 38 & 23.68\% & 7.89\% & 68.42\% & EP$>$(BT=CR) \\
     OT3 & 38 & 23.68\% & 7.89\% & 68.42\% & EP$>$(BT=CR) \\  
    \bottomrule
\end{tabular}
\end{table}

With regard to the \textit{technique participants applied best} (OT1), we can reject the null hypothesis that the frequency with which they perceive that they had applied each of the three techniques best is the same ($\chi^2$(2,N=38)=10.947, p=0.004). {\bf More people think they applied EP best, followed by both BT and CR} (which merit the same opinion). 

In the case of the \textit{technique participants liked best} (OT2), the results are similar. We can reject the null hypothesis that participants equally as often regard all three techniques as being their favourite technique ($\chi^2$(2,N=38)=22.474, p=0.000). {\bf Most people like EP best, followed by both BT and CR} (which merit the same opinion). 

Finally, as regards the \textit{technique that participants found easiest to apply} (OT3), the results are exactly the same as for the preferred technique ($\chi^2$(2,N=38)=22.474, p=0.000). {\bf Most people regard EP as being the technique that is easiest to apply, followed by both BT and CR} (which merit the same opinion). 

Table~\ref{tab:opip2} shows the results for participants’ opinions for programs. We cannot reject the null hypothesis of all programs equally frequently being viewed as \textit{the simplest}. ($\chi^2$(2,N=38)=1.474, p=0.479). Therefore, our data do not support the conclusion that {\bf all three programs are differently frequently perceived as being the simplest}. This result suggests that both the differences in complexity and size of cmdline and the highest Halstead metrics of ntree are small. This result suggest that participants could be interpreting differently this question. Another possibility could be that the question that has been used to operationalize the corresponding construct is vague, and participants are not interpreting it correctly.

\begin{table}[!hbt]
  \caption{Participants' Opinions for Programs in the Replicated Study}
  \label{tab:opip2}
  \begin{tabular}{lccccc}
    \toprule    
     Question & N & cmdline & nametbl & ntree  & Result \\
     \midrule      
     OP1 & 38 & 26.32\% & 42.11\% & 31.58\% & Cm=Na=Nt \\     
    \bottomrule
\end{tabular}
\end{table}

\subsubsection{RQ2.2: Comparing Opinions with Reality}

The technique that participants think they \textit{applied best} (OT1) {\bf is not a good predictor of technique effectiveness}. The overall and per technique kappa values in the fourth column of Table~\ref{tab:agreementorti2} are consistent with lack of agreement (in all cases ($\kappa<$0.4, poor), and although the upper bound of the 95\% CIs show agreement, 0 belongs to most 95\% CIs, meaning that agreement by chance cannot be ruled out). However, we find that there is a bias, as the Stuart-Maxwell and McNemar-Bowker tests can reject the null hypotheses of marginal homogeneity ($\chi^2$(2,N=38)=10.815, p=0.004) and symmetry ($\chi^2$(3, N=38)=12.067, p=0.007), respectively. Looking at the light and dark grey cells in the corresponding contingency table represented in Table~\ref{tab:contit2}, we find that the cells placed under the diagonal have higher values than those positioned above the diagonal. In other words, there are rather more participants that consider that they applied EP best, despite achieving better effectiveness with CR and BT (9 and 5), than participants who consider that they applied CR or BT best, despite being more effective using EP (1 in both cases). This suggests that {\bf there is a bias towards EP}. This bias is much more pronounced with respect to CR. These results are consistent with the ones found in the previous section. There are several possible interpretations for these results: 1) we do not know if the opinion on the best applied technique is accurate (meaning that it is really the best applied technique); 2) possibly due to the change in faults, technique performance is worse in this replication than in the original study; and 3) it could be that participants have misunderstood the question. Interviewing participants or asking them in the questionnaire about the reasons for their answers, would have helped to clarify this last issue.

\begin{table}[!hbt]
  \caption{Agreement between Opinions and Reality for Techniques in the Replicated Study}
  \label{tab:agreementorti2}
  \begin{tabular}{lllccc}
    \toprule    
      & & & & \multicolumn{2}{c}{95\% Confidence Interval} \\      \cmidrule{5-6}
      Question & N & Technique & Kappa value & Lower bound & Upper bound \\ \midrule  
          
 	  OT1 & 38 & Overall & 0.257 &  0.060 &  0.461 \\	
 	  & & CR & 0.348 &  0.091 & 0.598 \\	
 	  & & BT & 0.166 &  -0.181 & 0.539 \\	
 	  & & EP & 0.216 &  -0.022 & 0.474 \\ \midrule  

 	  OT2 & 38 & Overall & 0.166 & -0.023 &  0.360 \\	
 	  & & CR & 0.232 &  -0.009 &  0.486 \\	
 	  & & BT & 0.101 &  -0.145 & 0.457 \\	
 	  & & EP & 0.134 &  -0.072 &  0.365 \\ \midrule  
 	  
 	  OT3 & 38 & Overall & 0.166 &  -0.016 &  0.367 \\	
 	  & & CR & 0.232 & -0.019  & 0.491 \\	
 	  & & BT & 0.101 &  -0.145 &  0.469 \\	
 	  & & EP & 0.134 &  -0.077 &  0.351 \\	 	  
    \bottomrule
\end{tabular}
\end{table}

\begin{table}[!hbt]
  \caption{Contingency Table for Best Applied Technique (OT1) in the Replicated Study}
  \label{tab:contit2}
  \begin{tabular}{lcccc}
    \toprule    
     Best applied & \multicolumn{3}{c}{Reality} & \\ 
     \cmidrule{2-4}
     technique & CR & BT & EP & Total \\
    \midrule      
     CR & 9 & \cellcolor{mediumgray}0 & \cellcolor{lightgray}1 & 10 \\
     BT & \cellcolor{mediumgray}3 & 2 & \cellcolor{darkgray}1 & 6 \\
     EP & \cellcolor{lightgray}9 & \cellcolor{darkgray}5 & 8 & 22 \\
     \midrule    
     Total & 21 & 7 & 10 & 38 \\
    \bottomrule
\end{tabular}
\end{table}

As regards \textit{participants' favourite technique} (OT2), {\bf the results are similar}. This opinion does not predict technique effectiveness, since all kappa values in the fourth column of Table~\ref{tab:agreementorti2} denote lack of agreement (in all cases ($\kappa<$0.4, poor), and although the upper bound of the 95\% CIs show agreement, 0 belongs to all 95\% CIs, meaning that agreement by chance cannot be ruled out). Again, we find there is bias, as the Stuart-Maxwell  and the McNemar-Bowker tests can reject the null hypotheses of marginal homogeneity ($\chi^2$(2,N=38)=11.931, p=0.003) and symmetry ($\chi^2$(3,N=38)=11.974,p=0.007), respectively. Looking at the light and dark grey cells in Table~\ref{tab:contot23}, we again find {\bf that there is bias towards EP}. There are rather more participants that think that they applied EP better, despite being more effective using CR and BT (12 and 5), than participants that considered that they applied CR or BT better, despite being more effective using EP (1 in both cases). Note that the bias between CR and EP is more pronounced. Note that it is very unlikely that participants have not properly interpreted this question. It just seems that the technique they most like is not typically the most effective.

\begin{table}[!hbt]
  \caption{Contingency Table for Favourite (OT2) and Easiest to Apply (OT3) Technique in the Replicated Study}
  \label{tab:contot23}
  \begin{tabular}{lcccc}
    \toprule    
     Preferred / & \multicolumn{3}{c}{Reality} & \\ 
     \cmidrule{2-4}
     Easiest to apply & CR & BT & EP & Total \\
    \midrule      
     CR & 7 & \cellcolor{mediumgray}1 & \cellcolor{lightgray}1 & 9 \\
     BT & \cellcolor{mediumgray}1 & 1 & \cellcolor{darkgray}1 & 3 \\
     EP & \cellcolor{lightgray}12 & \cellcolor{darkgray}5 & 9 & 26 \\
     \midrule    
     Total & 20 & 7 & 11 & 38 \\
    \bottomrule
\end{tabular}
\end{table}

Finally, with respect to the \textit{technique that is easiest to apply} (OT3), {\bf we find that the results are exactly the same as for their preferred technique}. However, as we have seen in OT2, their preferred technique is not a good predictor of effectiveness (see third row of Table~\ref{tab:agreementorti2}), and there is bias towards EP (see light and dark grey cells in Table~\ref{tab:contot23}). These results are in line with a common claim in SE, namely, that {\bf developers should not base the decisions that they make on their opinions, as they are biased}. Again, it should be noted that participants might not be interpreting the question as we expected. Further research is necessary.

\begin{table}[!hbt]
  \caption{Agreement between Opinions and Reality for Programs in the Replicated Study (OP1, N=38)}
  \label{tab:agreementorpi2}
  \begin{tabular}{lccc}
    \toprule    
      & & \multicolumn{2}{c}{95\% Confidence Interval} \\      \cmidrule{3-4}
      & Kappa value & Lower bound & Upper bound \\ \midrule      
 	  Overall & 0.225 & -0.014 & 0.449 \\	
 	  cmdline & 0.066 & -0.213 & 0.371 \\	
 	  nametbl & 0.197 & -0.124 & 0.501 \\	
 	  ntree & 0.372 & 0.034 & 0.662 \\	
    \bottomrule
\end{tabular}
\end{table}

As far as the \textit{simplest program} is concerned, we find, as we did for the techniques, that {\bf it is not a good predictor of the program in which most defects were detected} (the values for overall and per program kappa in Table~\ref{tab:agreementorpi2} denote lack of agreement (in all cases ($\kappa<$0.4, poor), and although the upper bound of the 95\% CIs show agreement, 0 belongs to 95\% CIs---except ntree, meaning that agreement by chance cannot be ruled out). Unlike the opinions on techniques, {\bf we were not able to find any bias} this time, as neither the null hypothesis of marginal homogeneity ($\chi^2$(2,N=38)=1.621, p=0.445) nor symmetry ($\chi^2$(3,N=38)=3.286, p=0.350) can be rejected. This result suggests that {\bf  the programs that participants perceive to be the simplest are not necessarily the ones where most defects have been found}. Again, it should be noted the problem of participants interpreting the simplest program in a different way as we expected.

\subsubsection{Findings}

Our {\bf findings} suggest:

\begin{itemize}

\item Participants’ opinions should not drive their decisions. 

\item Participants prefer EP (think they applied it best, like it best and think that it is easier to apply), and rate CR and BT equally. 

\item All three programs are equally frequently perceived as being the simplest. 

\item The programs that the participants perceive as the simplest are not the ones where the highest number of defects have been found.

\end{itemize}

These results should be understood within the validity limits of the study.

\subsection{RQ3: Comparing Perceptions and Opinions}

\subsubsection{RQ3.1: Comparing Perceptions and Opinions}

In this section, we look at whether participants' perceptions of technique effectiveness are biased by their opinions about the techniques.

According to the results for kappa shown in the fourth column of Table~\ref{tab:agreementpoti2} (PT1-OT1), we find that {\bf results are compatible with agreement} (overall and per technique, except for BT in which lack of agreement cannot be ruled out) {\bf between the \textit{technique perceived to be the most effective} and the \textit{technique participants think they applied best}} (in all cases ($\kappa>$0.4, fair to good), and in all cases but BT, 0 does not belong to 95\% CIs, meaning that agreement by chance can be ruled out). This is an interesting finding, as it suggests that {\bf participants think that technique effectiveness is related to how well the technique is applied}. Technique performance definitely decreases if they are not applied properly. It is no less true, however, that techniques have intrinsic characteristics that may lead to some defects not being detected. In fact, the controlled experiment includes some faults that some techniques are unable to detect. A possible explanation for this result could be that the evaluation apprehension threat is materializing.

\begin{table}[!hbt]
  \caption{Agreement between Technique Perceptions and Opinions in the Replicated Study}
  \label{tab:agreementpoti2}
  \begin{tabular}{lllccc}
    \toprule    
      & & & & \multicolumn{2}{c}{95\% Confidence Interval} \\      \cmidrule{5-6}
      Question & N & Technique & Kappa value & Lower bound & Upper bound \\ \midrule  
          
 	  PT1-OT1 & 37 & Overall & \cellcolor{lightgray} 0.561 & 0.340 & 0.767 \\	
 	  & & CR & \cellcolor{lightgray} 0.513 & 0.216 & 0.801 \\	
 	  & & BT & \cellcolor{lightgray} 0.406 & -0.045 & 0.768 \\	
 	  & & EP & \cellcolor{lightgray} 0.684 & 0.479 & 0.891 \\ \midrule  

 	  PT1-OT2 & 37 & Overall & 0.325 & 0.100 & 0.541 \\	
 	  & & CR & 0.197 & -0.121 & 0.510 \\	
 	  & & BT & 0.323 & -0.078 & 0.684 \\	
 	  & & EP & \cellcolor{lightgray} 0.432 & 0.180 & 0.682 \\ \midrule  
 	  
 	  PT1-OT3 & 37 & Overall & 0.325 & 0.103 & 0.543 \\	
 	  & & CR & 0.197 & -0.117 & 0.508 \\	
 	  & & BT & 0.323 & -0.088 & 0.684 \\	
 	  & & EP & \cellcolor{lightgray} 0.432 & 0.174 & 0.674 \\	 	  
    \bottomrule
\end{tabular}
\end{table}

On the other hand, the kappa values in the fourth column of Table~\ref{tab:agreementpoti2} (PT1-OT2) \textbf{reveal a lack of agreement for CR and BT between the \textit{preferred technique} and the \textit{technique perceived as being most effective}} (in all cases ($\kappa<$0.4, poor), and although the upper bound of the 95\% CIs show agreement, 0 belongs to 95\% CIs, meaning that agreement by chance cannot be ruled out),  \textbf{whereas overall lack of agreement cannot be ruled out} (($\kappa<$0.4, poor), the upper bound of the 95\% CI shows agreement, and 0 does not belong the to 95\% CI, meaning that agreement by chance can be ruled out). Finally, \textbf{there is agreement} ($\kappa>$0.4, fair to good) \textbf{in the case of EP}. This means that, {\bf in the case of EP, participants tend to associate their favourite technique with the perceived most effective technique, contrary to the findings for CR and BT}. This is more likely to be due to EP being the technique that many more participants like best (and the chances for there being a match are higher compared to the other techniques) than to there actually being a real match.

With respect to directionality whenever there is disagreement, the results of the Stuart-Maxwell and the McNemar-Bowker tests show that the null hypotheses of marginal homogeneity ($\chi^2$(2,N=37)=8.355,p=0.015) and symmetry ($\chi^2$(3,N=37)=8.444, p=0.038) can be rejected. Looking at the light grey cells in Table~\ref{tab:contit2ot23}, we find that there are more participants claiming to have applied CR best that prefer EP than vice versa (8 versus 1). This means that {\bf the mismatch between the technique that participants like best and the technique that they perceive as being most effective can largely be put down to participants who like EP better perceiving CR to be more effective}.

\begin{table}[!hbt]
  \caption{Contingency Table for Perceived Most Effective vs. Preferred Technique (PT1-OT2) and Perceived Most Effective vs. Easiest to Apply Technique (PT1-OT3)}
  \label{tab:contit2ot23}
  \begin{tabular}{lp{0.9cm}p{0.9cm}p{0.9cm}c}
    \toprule    
     Perceived & \multicolumn{3}{c}{Preferred/Easiest to apply} & \\ 
     \cmidrule{2-4}
     most effective & CR & BT & EP & Total \\
    \midrule      
     CR & 5 & \cellcolor{mediumgray}1 & \cellcolor{lightgray}8 & 14 \\
     BT & \cellcolor{mediumgray}3 & 2 & \cellcolor{darkgray}2 & 7 \\
     EP & \cellcolor{lightgray}1 & \cellcolor{darkgray}0 & 15 & 16 \\
     \midrule    
     Total & 9 & 3 & 25 & 37 \\
    \bottomrule
\end{tabular}
\end{table}

{\bf The results for the agreement between the \textit{technique that is easiest to apply} and the \textit{technique that is perceived to be most effective}} are exactly the same as for the preferred technique (see third row of Table~\ref{tab:agreementpoti2}). This means that, for EP, the participants equate the technique that they find easiest to apply with the one that they regard as being most effective. This does not hold for the other two techniques. Likewise, the mismatch between the technique that is easiest to apply and the technique perceived as being most effective can be largely put down to participants who applied EP best perceiving CR to be more effective (see Table~\ref{tab:contit2ot23}). 

As mentioned earlier, we found that participants have a correct perception of the program in which they detected most defects. Table~\ref{tab:agreementpopi2} shows that participants do not associate \textit{simplest program} with \textit{program in which most defects were detected} (PP1-OP1). This is striking as it would be logical for it to be easier to find defects in the simplest program. As illustrated by the fact that the null hypotheses of marginal homogeneity ($\chi^2$(2,N=37)=3.220,p=0.200) and symmetry ($\chi^2$(3,N=37)=4.000, p=0.261) cannot be rejected, we were not able to find bias in any of the cases where there is disagreement. A possible explanation for this result is that participants are not properly interpreting what simple means.

\begin{table}[!hbt]
\caption{Agreement between Program Perceptions and Opinions in the Replicated Study (PP1-OP1, N=37)}
  \label{tab:agreementpopi2}
  \begin{tabular}{lccc}
    \toprule    
      & & \multicolumn{2}{c}{95\% Confidence Interval} \\      \cmidrule{3-4}
      & Kappa value & Lower bound & Upper bound \\ \midrule      
 	  Overall & 0.189 & -0.057 &  0.433 \\	
 	  cmdline & 0.308 & -0.029 & 0.595 \\	
 	  nametbl & 0.043 & -0.265 & 0.359 \\	
 	  ntree & 0.228 & -0.097 & 0.590 \\	
    \bottomrule
\end{tabular}
\end{table}

\subsubsection{RQ3.2: Comparing Opinions}

Finally, we study the possible relation between the opinions themselves. Looking at Table~\ref{tab:agreementooi2}, we find that {\bf participants equate \textit{the technique they applied best} with \textit{their favourite technique} and with \textit{the technique they found easiest to apply}} (overall and per technique ($\kappa>$0.4, fair to good), and 0 does not belong to 95\% CIs, meaning that agreement by chance can be ruled out). It makes sense that the technique that participants found easiest to apply should be the one that they think they applied best and like best. Typically, people like easy things (or maybe we think things are easy because we like them). In this respect, we can conclude that participants’ opinions about the techniques all have the same directional effect.

\begin{table}[!hbt]
  \caption{Agreement among Technique Opinions in the Replicated Study}
  \label{tab:agreementooi2}
  \begin{tabular}{lllccc}
    \toprule    
      & & & & \multicolumn{2}{c}{95\% Confidence Interval} \\      \cmidrule{5-6}
      Question & N & Technique & Kappa value & Lower bound & Upper bound \\ \midrule  
          
 	  OT1-OT2 & 38 & Overall & \cellcolor{lightgray} 0.652 & 0.411 & 0.858 \\	
 	  & & CR & \cellcolor{lightgray} 0.649 & 0.350 & 0.887 \\	
 	  & & BT & \cellcolor{lightgray} 0.627 & 0.001 & 1.000 \\	
 	  & & EP & \cellcolor{lightgray} 0.665 & 0.410 & 0.892 \\ \midrule  

 	  OT1-OT3 & 38 & Overall & \cellcolor{lightgray} 0.652 & 0.405 & 0.865 \\	
 	  & & CR & \cellcolor{lightgray} 0.649 & 0.319 & 0.907 \\	
 	  & & BT & \cellcolor{lightgray} 0.627 & 0.000 & 0.907 \\	
 	  & & EP & \cellcolor{lightgray} 0.665 & 0.408 & 0.887 \\ \midrule  
 	  
 	  OT2-OT3 & 38 & Overall & \cellcolor{lightgray} 1.000 & 1.000 &  1.000 \\	
 	  & & CR & \cellcolor{lightgray} 1.000 & 1.000 & 1.000 \\	
 	  & & BT & \cellcolor{lightgray} 1.000 & 1.000 & 1.000 \\	
 	  & & EP & \cellcolor{lightgray} 1.000 & 1.000 & 1.000 \\	 	  
    \bottomrule
\end{tabular}

\end{table}

\subsubsection{Findings}

Our {\bf findings} suggest:

\begin{itemize}

\item Participants’ perceptions of technique effectiveness are related to how well they think they applied the techniques. They tend to think it is they, rather than the techniques, that are the obstacle to achieving more effectiveness (a possible evaluation apprehension threat has materialized). 

\item We have not been able to find a relationship between the technique they like best and find easiest to apply, and perceived effectiveness. Note however, that the technique participants think they have applied best is not necessarily the one that they have really best applied.

\item Participants do not associate the simplest program with the program in which they detected most defects. This could be due to participants not properly interpreting the concept "simple".

\item Opinions are consistent with each other.

\end{itemize}

Again, these results are confined to the validity limits imposed by the study.

\section{Discussion}
\label{sec:discussion}

Next, we summarize the findings of this study and analyse their implications. Note that the results of the study are restricted to junior programmers with little testing experience, and defect detection techniques.

\subsection{Answers to Research Questions}

\begin{itemize}

    \item \textit{RQ1.1: What are participants' perceptions of their testing effectiveness?} 
   
   The number of participants perceiving a particular technique/program as being more effective cannot be considered different for all three techniques/programs.  
   
    \item \textit{RQ1.2: Do participants' perceptions predict their testing effectiveness?} 
    
    Our data do not support that participants correctly perceive the most effective technique for them. Additionally, no bias has been found towards a given technique. However, they tend to correctly perceive the program in which they detected most defects.

    \item \textit{RQ1.3: Do participants find a similar amount of defects for all techniques?}
   	
   	Participants do not obtain similar effectiveness values when applying the different techniques.
   	
    \item \textit{RQ1.4: What is the cost of any mismatch?} 
    
    Mismatch cost is not negligible (mean 31pp), and it is not related to the technique perceived as most effective.
   
    \item \textit{RQ1.5: What is expected project loss?} 

	Expected project loss is 15pp, and it is not related to the technique perceived as most effective.
    
    \item \textit{RQ1.6: Are participants perceptions related to the number of defects reported by participants?}

	Results are not clear about this. Although our data do not support that participants correctly perceive the most effective technique for them, it should not be ruled out. Further research is needed.

\end{itemize}

Therefore, the answer to \textit{RQ1: Should participants' perceptions be used as predictors of testing effectiveness?} is that participants should not base their decisions on their own perceptions, as they are not reliable and have an associated cost.

\begin{itemize}

    \item \textit{RQ2.1: What are participants' opinions about techniques and programs?} 
 
    Most people like EP best, followed by both BT and CR (which merit the same opinion). There is no difference in opinion as regards programs

    \item \textit{RQ2.2: Do participants' opinions predict their effectiveness?} 
   
    They are not good predictors of technique effectiveness. A bias has been found towards EP.

    \end{itemize}

Therefore, the answer to \textit{RQ2: Can participants' opinions be used as predictors for testing effectiveness?} is that participants should not use their opinions, as they are not reliable and are biased.

\begin{itemize}

    \item \textit{RQ3.1: Is there a relationship between participants' perceptions and opinions?}

Participants' perceptions of technique effectiveness are related to how well they think they applied the techniques. We have not been able to find a relationship between the technique they like best and find easiest to apply, and perceived effectiveness. Participants do not associate the simplest program with the program in which they detected most defect.

    \item \textit{RQ3.2: Is there a relationship between participants' opinions?}
   
    Yes. Opinions are consistent with each other.

\end{itemize}

Therefore, the answer to \textit{RQ3: Is there a relationship between participants' perceptions and opinions?} is positive for some of them.

\subsection{About Perceptions}
Participants’ perceptions about the effectiveness of techniques are incorrect (50\% get it wrong). However, this is not due to some sort of bias in favour of any of the three techniques under review. These misperceptions should not be overlooked, as they affect software quality. We cannot accurately estimate the cost, as it depends on what faults there are in the software. However, our data suggest a loss of from 25pp to 31 pp. Perceptions about programs appear to be correct, although this does not offset the mismatch cost.

Our findings {\bf confirm} that:

\begin{itemize}

\item Testing technique effectiveness depends on the software faults.

\end{itemize}

Additionally, they {\bf warn} developers that:

\begin{itemize}

\item They should not rely on their perceptions when rating a defect detection technique or how well they have tested a program.

\end{itemize}

Finally, they suggest the need for the following {\bf actions}:

\begin{itemize}

\item Develop tools to inform developers about how effective the techniques that they applied are and the testing they performed is.

\item Develop instruments to give developers access to experimental results. 

\item Conduct further empirical studies to learn what technique or combination of techniques should be applied under which circumstances to maximize its effectiveness.

\end{itemize}

\subsection{About Opinions}

Participants prefer EP to BT and CR (they like it better, think they applied it better and find it easier to apply). Opinions do not predict real effectiveness. This failure to predict reality is partly related to the fact that a lot of people prefer EP but are really more effective using BT or CR. Opinions do not predict real effectiveness with respect to programs either.

These findings {\bf warn} developers that:

\begin{itemize}

\item They should not be led by their opinions on techniques when rating their effectiveness.

\end{itemize}

Finally, they suggest the need for the {\bf action}: 

\begin{itemize}

\item Further research should be conducted into what is behind developers' opinions.

\end{itemize}

\subsection{About Perceptions and Opinions}

The technique that participants believe to be the most effective is the one that they applied best. However, they are capable of separating their opinions about technique complexity and preferences from their perceptions, as the technique that they think is most effective is not the one that they find easiest to apply or like best.

Our findings {\bf challenge} that:

\begin{itemize}

\item Perceptions of technique effectiveness are based on participants’ preferences.

\end{itemize}

They also {\bf warn} developers that:

\begin{itemize}

\item Maximum effectiveness is not necessarily achieved when a technique is properly applied.

\end{itemize}

Finally, they suggest the need for the following {\bf actions}:

\begin{itemize}

\item Determine the best combination of techniques to apply that is at the same time easily applicable and effective.

\item Continue to look for possible drivers to determine what could be causing developers' misperceptions.

\end{itemize}


\section{Related Work}
\label{sec:related_work}

In recent years, several \textit{experiments on defect detection technique effectiveness (static techniques and/or test-case design techniques) have been run with and without humans}. Experiments without human compare the efficiency and effectiveness of specification-based, code-based, and fault-based techniques, as for example the ones conducted by Bieman \& Schultz \cite{BS:92}, Hutchins \textit{et al.} \cite{HFGO:94}, Offut \textit{et al.} \cite{OLRUZ:96}, Offut \& Lee \cite{OL:94}, Weyuker \cite{W:84} and Wong \& Mathur \cite{WM:95}. Most of the experiments with humans evaluate static techniques, as for example the ones run by Basili \textit{et al.} \cite{BGLLSSZ:96}, Biffl \cite{B:00}, Dunsmore \textit{et al.} \cite{DRW:02}, Maldonado \textit{et al.}\cite{MCSFDMMB:06}, Porter \textit{et al.} \cite{PVB:95} and Thelin \textit{et al.} \cite{TRWOA:04}. Experiments evaluating test-case design techniques studied the efficiency and effectiveness of specification-based and control-flow-code-based techniques applied by humans, as the ones run by Basili \& Selby \cite{BS:87}, Briand \textit{et al.} \cite{BPL:04}, Kamsties \& Lott \cite{KL:95}, Myers \cite{M:78} and Roper \textit{et al.} \cite{RWM:97}. These experiments focus on strictly quantitative issues, leaving aside human factors like developers' perceptions and opinions.

There are \textit{surveys that study developers' perceptions and opinions} with respect to different testing issues, like the ones performed by Deak \cite{D:12}, Dias-Neto \textit{et al.}\cite{DMSRT:16}, Garousi \textit{et al.} \cite{GFKH:17}, Goncalves \textit{et al.} \cite{GAAFXF:17}, Guaiani \& Muccini \cite{GM:15}, Khan \textit{et al.} \cite{KPCA:10} and Hernández \& Marsden \cite{PRHM:14}. However, the results are not linked to quantitative issues. In this regard, some studies \textit{link personality traits to preferences according to the role of software testers}, as for example Capretz \textit{et al.} \cite{CVR:15}, Kanij \textit{et al.} \cite{KMG:15} and Kosti \textit{et al.} \cite{KFA:14}. However, there are no studies looking for a relationship between personality traits and quantitative issues like testing effectiveness. 

There are some \textit{approaches for helping developers to select the best testing techniques to apply under particular circumstances}, like the ones made by Cotroneo \textit{et al.}\cite{CPR:13}, Dias-Neto \& Travassos \cite{DT:14} or Vegas \textit{et al.} \cite{VJB:09}. Our study suggests that this type of research needs to be more widely disseminated to improve knowledge about techniques.

Finally, there are several ways in which developers can make decisions in the software deveelopment industry. The most basic approach is the classical perceptions and/or opinions, as reported in Dybå \textit{et al.} \cite{DKJ:05} and Zelkowitz \textit{et al.} \cite{ZWB:03}. Other approaches suggest using classical decision-making models \cite{AW:16}. Experiments can also be used for industry decision-making, as described by Jedlitschka \textit{et al.} \cite{JJR:14}. Devanbu \textit{et al.} \cite{DZB:16} have observed the use of past experience (beliefs). More recent approaches advocate automatic decision-making from mining repositories\cite{B:12}.


\section{Conclusions}
\label{sec:conclusions}

The goal of this paper was to discover whether developers' perceptions of the effectiveness of different code evaluation techniques are right in absence of prior experience. To do this, we conducted an empirical study with students plus a replication. The original study revealed that participants’ perceptions are wrong. As a result, we conducted a replication aimed at discovering what was behind participants’ misperceptions. We opted to study participants’ opinions on techniques. The results of the replicated study corroborate the findings of the original study. They also reveal that participants’ perceptions of technique effectiveness are based on how well they applied the techniques. We also found that participants’ perceptions are not influenced by their opinions about technique complexity and preferences for techniques.

Based on these results, we derived some recommendations for developers: they should not trust their perceptions and be aware that correct technique application does not assure that they will find all the program defects.

Additionally, we identified a number of lines of action that could help to mitigate the problem of misperception, such as developing tools to inform developers about how effective their testing is, conducting more empirical studies to discover technique applicability conditions, developing instruments to allow easy access to experimental results, investigating other possible drivers of misperceptions or investigating what is behind opinions. 

Future work includes running new replications of these studies to better understand their results.


\bibliographystyle{spmpsci}      
\bibliography{biblio}   

%
%

\clearpage

\hfill \break
\begin{floatingfigure}{1.5in}
\includegraphics[width=1in,clip,keepaspectratio]{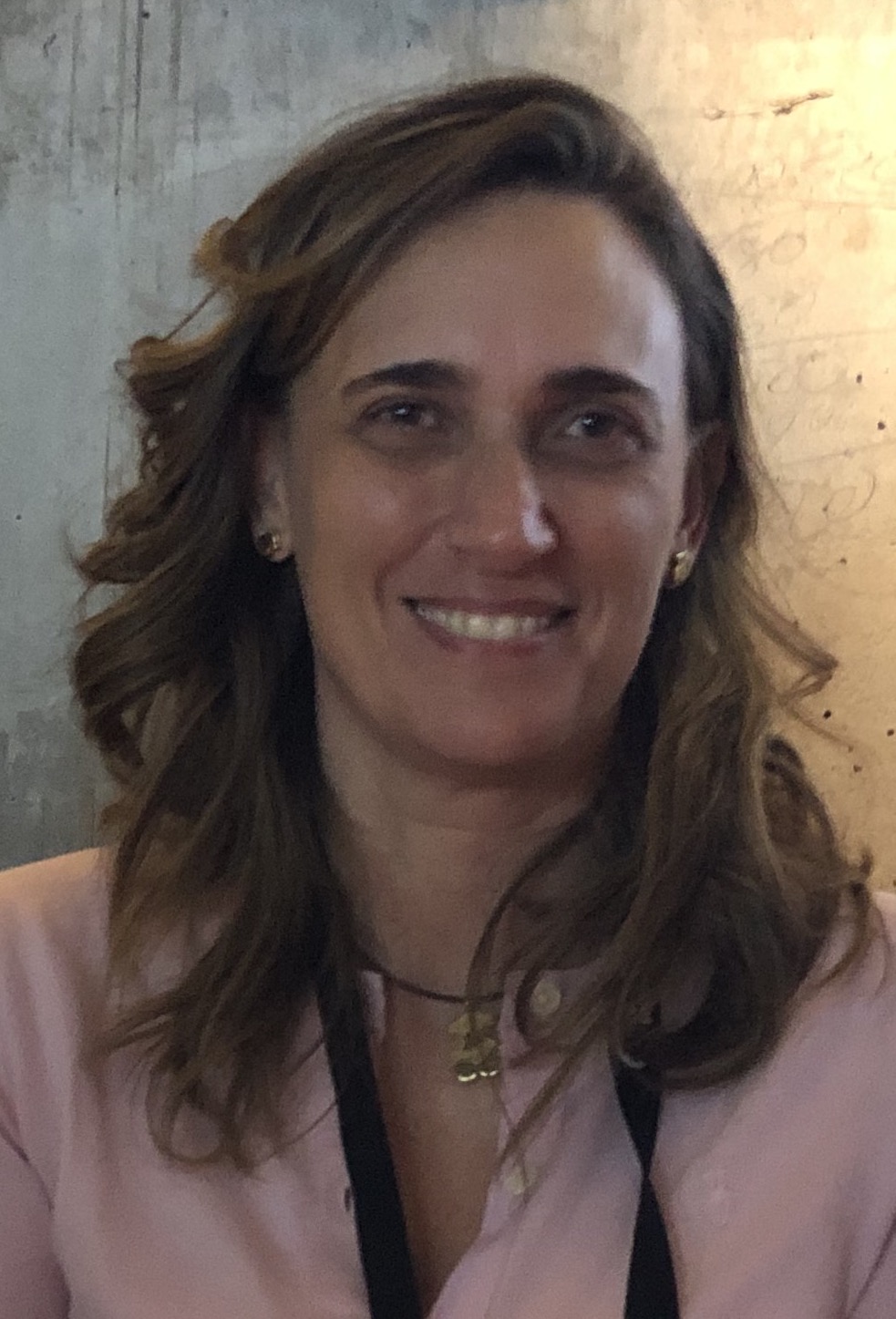}
\end{floatingfigure}
\noindent \textbf{Sira Vegas} has been associate professor of software engineering with the  School of Computer Engineering at the Technical University of Madrid, Spain, since 2008. Sira belongs to the review board of IEEE Transactions on Software Engineering, and is a regular reviewer of the Empirical Software Engineering Journal. She was program chair for the International Symposium on Empirical Software Engineering and Measurement in 2007.
\newline

 
\hfill 
\begin{floatingfigure}{1.5in}
\includegraphics[width=1in,clip,keepaspectratio]{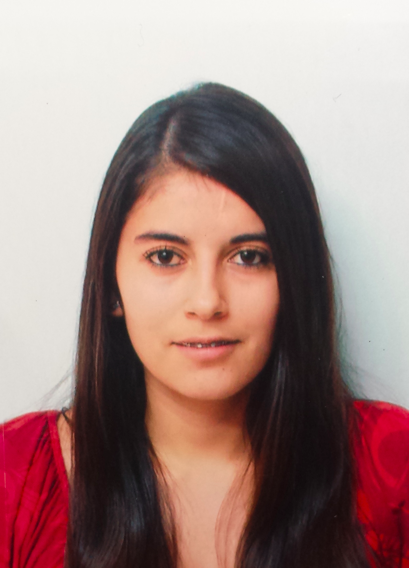}
\end{floatingfigure}
\noindent \textbf{Patricia Riofrío} received her BSc in Computer Science and her MSc in Software and Systems from the Polytechnic University of Madrid, Spain. She is a PhD student at the Polytechnic University of Madrid. Her main research interest is empirical software engineering. 
\newline
 
\vspace{1cm}

\hfill 
 \begin{floatingfigure}{1.5in}
 \includegraphics[width=1.3in,clip,keepaspectratio]{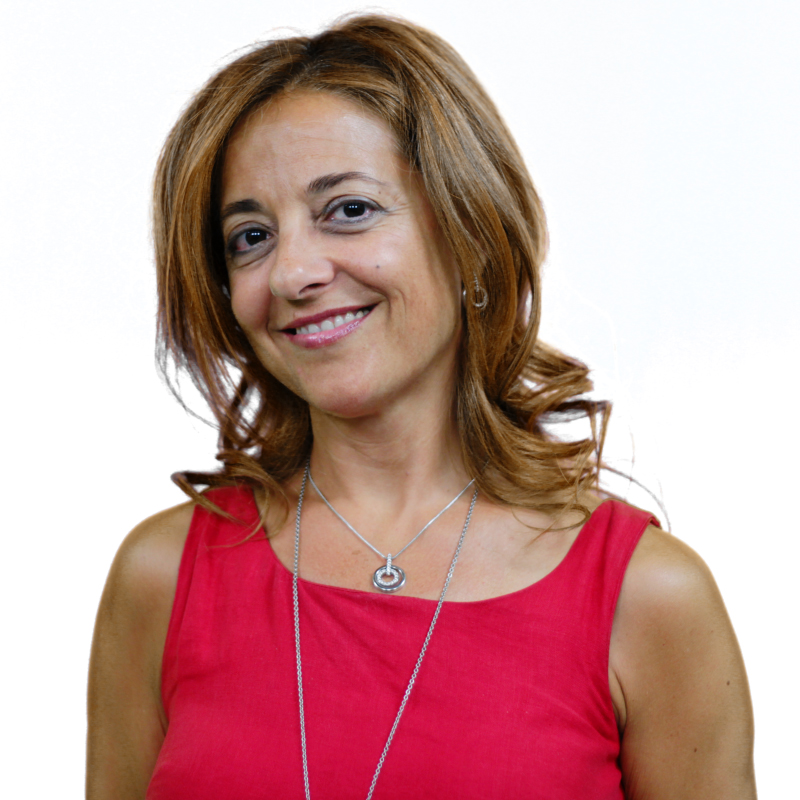}
 \end{floatingfigure}
 \noindent \textbf{Esperanza Marcos} is a full professor at the Universidad Rey Juan Carlos of Madrid, where she leads the Kybele R\&D group. She received her Ph.D. from the Technical University of Madrid in 1997 and her current research interests include Service Science, Model-Driven engineering and human aspects of Software Engineering. 
\newline

\vspace{1cm}
    
\hfill     
 \begin{floatingfigure}{1.5in}
 \includegraphics[width=1in,clip,keepaspectratio]{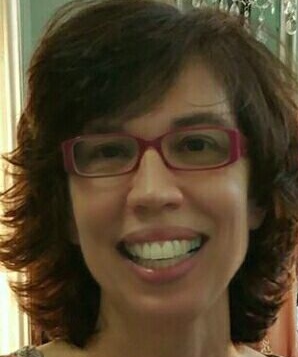}
 \end{floatingfigure}
 \noindent \textbf{Natalia Juristo} has been full professor of software engineering with the  School of Computer Engineering at the Technical University of Madrid, Spain, since 1997. She was awarded a FiDiPro (Finland Distinguished Professor Program) professorship at the University of Oulu, from January 2013 until June 2018. Natalia belongs to the editorial board of EMSE and STVR. In 2009, Natalia was awarded an honorary doctorate by Blekinge Institute of Technology in Sweden.

\begin{appendices}

\section{Program Metrics}
\label{app:metrics_programs}

Table~\ref{tab:metrics_programs} shows the metrics collected for each program with the PREST tool. Note that all three programs show similar results for all metrics, except ntree that shows higher Halstead metrics. The size and complexity of cmdline is slightly higher compared to the other two programs.

\begin{table}[!hbt]
  \caption{Metrics Obtained with PREST}
  \label{tab:metrics_programs}
  \begin{tabular}{lccc}
    \toprule    

Metric & cmdline & nametbl & ntree \\  \midrule      
Total loc &	289	& 289	& 247 \\
Blank LOC &	44 & 54	& 27 \\
Comment LOC & 6	& 5	& 5 \\
Code and Comment LOC &	0 & 0 & 0 \\
Executable LOC &	239	& 230	& 215 \\
Unique Operands &	91	& 121 &	89 \\
Unique Operators &	29 &	19 &	21 \\
Total Operands &	401	& 486	& 609 \\
Total Operators &	560	& 594	& 698 \\
Halstead Vocabulary &	120	& 140	& 110 \\
Halstead Length &	961	& 1080	& 1307 \\
Halstead Volume &	4600 &	5336 &	6143 \\
Halsted Level &	0.02	& 0.03 &	0.01 \\
Halstead Difficulty &	50.0 &	33.33 &	100.0 \\
Halstead Effort 	& 230000.0 &	177866.67 &	614300.0 \\
Halstead Error 	& 1.53 &	1.78 &	2.05 \\
Halstead Time 	& 12777.78 &	9881.48	& 34127.78 \\
Branch Count &	108	& 84	& 94 \\
Decision Count 	& 54	& 42	& 47 \\
Call Pairs &	41 &	57 &	35 \\
Condition Count &	51	& 40	& 39 \\
Multiple Condition Count &	16 &	16 &	 17 \\
Cyclomatic Complexity &	37 &	27 & 	31 \\
Cyclomatic Density & 	0.15 &	0.12	 & 0.14 \\
Decision Density &	1.06 &	1.05 &	1.21 \\
Design Complexity &	41 &	57 &	35 \\
Design Density &	1.11 &	2.11 &	1.13 \\
Normalized Cyclomatic Complexity & 	0.13	 & 0.09 &	0.13 \\
Formal Parameteres &	0	& 0 &	0 \\
Risk Level	& False	& False	& False \\

     \bottomrule  
\end{tabular}
\end{table}

\section{Analysis of the Original Experiment}
\label{app:experiment_it1}

\begin{figure}[!hbt]
\includegraphics[width=3.5in]{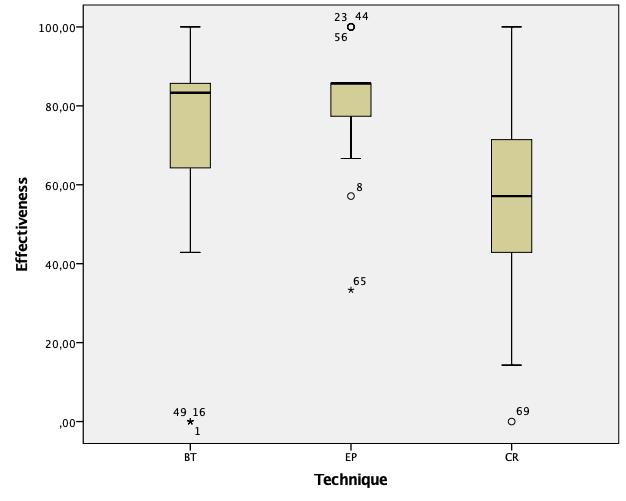}
\caption{Boxplot for observed technique effectiveness in the original study}
\label{fig:bpti1}
\end{figure}

Figure~\ref{fig:bpti1} shows the boxplot, and Table~\ref{tab:dsti1} shows the descriptive statistics for observed technique effectiveness.

\begin{table}[!hbt]
  \caption{Descriptive Statistics for Observed Technique Effectiveness in the Original Study}
  \label{tab:dsti1}
  \begin{tabular}{lcccc}
    \toprule    
     & & Std. & \multicolumn{2}{c}{95\% Confidence Interval} \\ 
     \cmidrule{4-5}
     Technique & Mean & Deviation & Lower Bound & Upper Bound \\    
    \midrule      
    CR & 52.795 & 24.904 & 42.025 & 63.564 \\
    BT & 66.251 & 33.691 & 51.682 & 80.820 \\
    EP & 82.089 & 15.921 & 75.204 & 88.973 \\

    \bottomrule
\end{tabular}
\end{table}

We find that the mean and median effectiveness of BT is highest, followed by EP and then by CR. Additionally, EP has a lower variance. The 95\% confidence interval suggests that EP is more effective than CR and that BT is as effective as EP and CR. This is an interesting result, as all faults injected in the code could be detected by all techniques. Additionally, it could indicate that code reading is more dependent on experience that you cannot acquire in a 4-hour training.

All three techniques have outliers corresponding to participants that have performed exceptionally bad: 3 in the case of BT (all 3 participants scored 0), 2 in the case of EP and 1 in the case of CR (also scoring 0). Additionally, EP has 3 outliers corresponding to participants that have performed exceptionally well (all scoring 100). In none of the cases the values belong to the same participant (which could suggest that outliers correspond to participants that performed exceptionally bad with one technique, but not all three). Additionally, CR shows a higher variability, which could indicate that it is more dependent on the person applying the technique.

The experimental data has been analysed with a Linear Mixed-Effects Model (SPSS v26 MIXED procedure). Group, program, technique and the program by technique interaction are fixed effects and subject is a random effect. Eleven models were tried, choosing the one with the lowest AIC:

\begin{itemize}
	\item One pure random effects model and no repeated measures.
	\item Five models specifying program as repeated measures effect.
	\item Five models specifying technique as repeated measures effect.
\end{itemize}

The five models differed in the covariance structures used (identity, diagonal, first-order autoregressive, compound symmetry and unstructured).

In this first analyses the MIXED procedure did not achieve convergence. We then decided to relax the model by removing subject from the random effects list. We re-run analyses, but this time we found severe departures from normality, leading to non-reliable results. Next step consisted on data transformation. The chosen transformation, square, solved the normality issues, and the MIXED procedure converged. 

\begin{figure}[!hbt]
\includegraphics[width=4in]{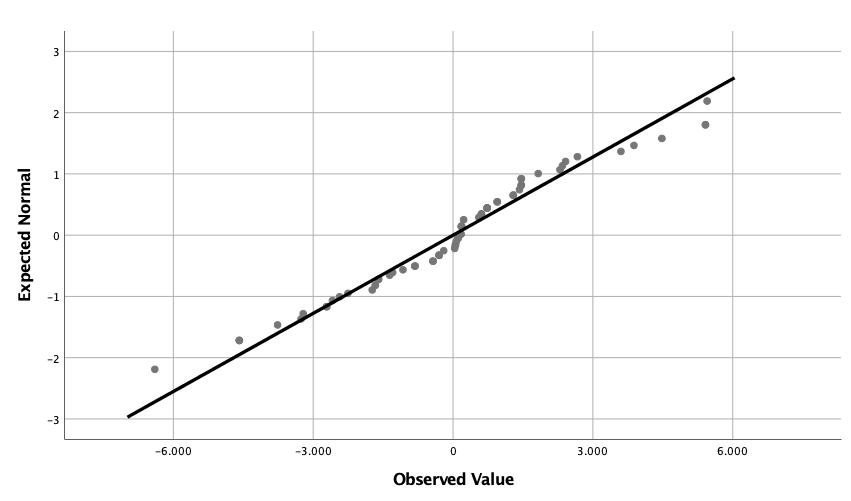}
\caption{Normal Q-Q plot of residuals in the original study}
\label{fig:qq_it1}
\end{figure}

Table~\ref{tab:mixed_model_it1} shows the results for the model chosen (program is a repeated measures effect, and the covariance structure is Diagonal). Figure~\ref{fig:qq_it1} shows residuals normality.

\begin{table}[!hbt]
  \caption{Type III Tests of Fixed Effects in the Original Study}
  \label{tab:mixed_model_it1}
  \begin{tabular}{lcccc}
    \toprule     

	Source & Numerator df & Denominator df & F & Sig. \\
     \midrule      
 	Intercept &  1 & 57.182 & 338.655 & 0.000 \\
 	Group & 5 & 45.470 & 0.259 & 0.933 \\
 	Technique & 2 & 54.522 & 16.648 &  0.000 \\
 	Program & 2 & 51.485 & 6.852 & 0.002 \\
 	Technique * Program & 4 & 61.073 & 0.583 & 0.676 \\
  
     \bottomrule  
\end{tabular}
\end{table}

Table~\ref{tab:mixed_model_it1} shows that technique and program are both statistically significant. Group and the technique by program interaction are not significant. The Bonferroni multiple comparisons tests show that:

\begin{itemize}
	\item All three techniques show different effectiveness (EP$>$BT$>$CR).
	\item ntree shows a higher defect detection rate than the other two programs (ntree$>$(cmdline$=$nametbl)).
\end{itemize} 

Table~\ref{tab:technique_means_it1} and Table~\ref{tab:program_means_it1} show the estimated marginal means for both technique and program. Note that the mean, std. error and 95\% confidence interval bounds values have been un-transformed.

\begin{table}[!hbt]
  \caption{Estimated Marginal Means for Technique in the Original Study}
  \label{tab:technique_means_it1}
  \begin{tabular}{lccccc}
    \toprule     

	 &  & & & \multicolumn{2}{c}{95\% Confidence Interval} \\
	 \cmidrule{5-6}
	Technique & Mean & Std. Error & df & Lower Bound & Upper Bound \\
     \midrule      
 	BT & 74.307	& 3.309 & 53.279 & 67.199 &	80.793 \\
 	EP &  85.486 & 2.864 & 57.329 & 79.477 & 91.099  \\
 	CR &  57.502 & 4.118 & 53.212 & 48.096 & 65.572   \\
 	  
     \bottomrule  
\end{tabular}
\end{table}

\begin{table}[!hbt]
  \caption{Estimated Marginal Means for Program in the Original Study}
  \label{tab:program_means_it1}
  \begin{tabular}{lccccc}
    \toprule     

	 &  & & & \multicolumn{2}{c}{95\% Confidence Interval} \\
	 \cmidrule{5-6}
	Program & Mean & Std. Error & df & Lower Bound & Upper Bound \\
     \midrule      
 	cmdline & 66.562 & 4.655 & 21.854 & 55.969 & 75.686  \\
 	nametbl & 69.642 & 2.271 & 23.153 & 64.886 & 74.093  \\
 	ntree &  82.798	& 3.117 & 21.900 & 76.201 & 88.907  \\
  
     \bottomrule  
\end{tabular}
\end{table}

\begin{figure}[!hbt]
\includegraphics[width=4.5in]{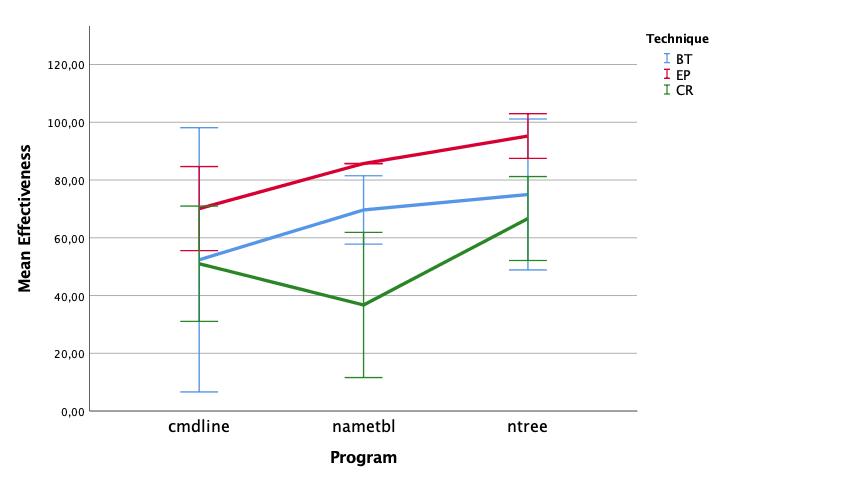}
\caption{Profile plot for program by technique interaction in the original study}
\label{fig:pp_pt_it1}
\end{figure}

Finally, Figure~\ref{fig:pp_pt_it1} shows the profile plot with error bars for the program by technique interaction. Although the interaction is not statistically significant, the profile plot suggests that nametbl could be behaving differently for CR. This could mean that lack of significance of the interaction in the analysis could be due to sample size.

\section{Analysis of the Replicated Experiment}
\label{app:experiment_it2}

\begin{figure}[!hbt]
\includegraphics[width=3.5in]{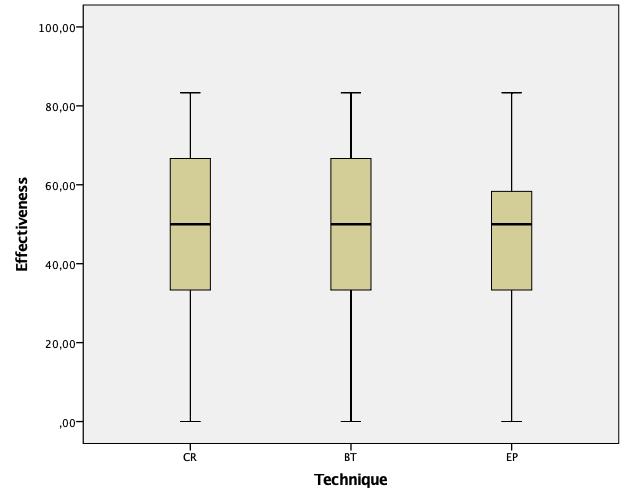}
\caption{Boxplot for observed technique effectiveness in the replicated study.}
\label{fig:bpti2}
\end{figure}

Figure~\ref{fig:bpti2} shows the boxplot for observed technique effectiveness. All three techniques show a similar median, and the same range. Compared to the original study, the behaviour of the techniques is much more homogeneous.

Table~\ref{tab:dsti2} shows the descriptive statistics for observed technique effectiveness. We find that the effectiveness of all three techniques is similar. CR has a similar effectiveness as in the original study suggesting again that perhaps 4 hours of training are not enough for learning the code review technique). However, the mean effectiveness for BT and EP has dropped from 66\% and 82\%, respectively, in the original study to 46\%. Note that in this study, the nature of the faults has been changed. While in the original study all defects can be detected by all techniques, in this study some defects cannot be exercised by testing techniques. Therefore, a possible explanation for the change in effectiveness of testing techniques could be the faults seeded in the programs. However, this is just a mere hypothesis that need to be tested.

\begin{table}[!hbt]
  \caption{Descriptive Statistics for Observed Technique Effectiveness in the Replicated Study}
  \label{tab:dsti2}
  \begin{tabular}{lcccc}
    \toprule    
     & & Std. & \multicolumn{2}{c}{95\% Confidence Interval} \\ 
     \cmidrule{4-5}
     Technique & Mean & Deviation & Lower Bound & Upper Bound \\    
    \midrule      
     CR & 52.564 & 22.793 & 45.175 & 59.953 \\
     BT & 46.153 & 17.297 & 40.546 & 51.760 \\
     EP & 46.581 & 17.597 & 40.876 & 52.285 \\
    \bottomrule
\end{tabular}
\end{table}

\begin{figure}[!hbt]
\includegraphics[width=3.5in]{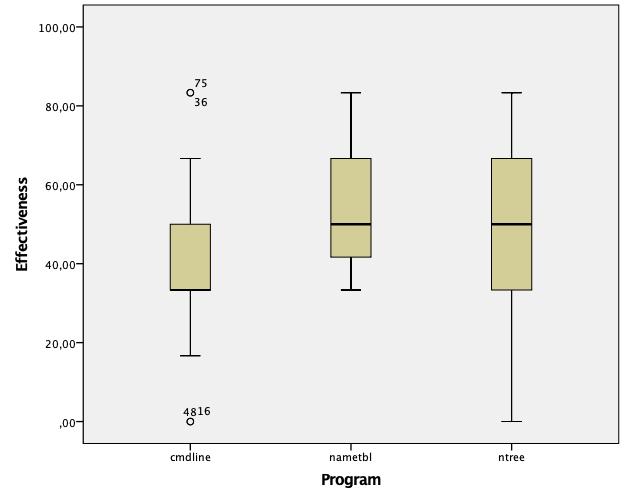}
\caption{Boxplot for observed program detection rate in the Replicated Study}
\label{fig:bppi2}
\end{figure}

Figure~\ref{fig:bppi2} shows the boxplot for the percentage of defects found in each program. It is interesting to see that the median detection rate for nametbl and ntree is higher than for cmdline. These results could be attributed to cmdline having a slightly higher size and complexity. Additionally, cmdline shows 4 outliers, reflecting 4 people who performed exceptionally well, and 2 people who performed exceptionally bad (scoring 0). Finally, ntree shows a higher range compared to the other two programs (note that ntree shows higher Halstead metrics). This is an unexpected result, as nametbl and ntree are very similar in terms of complexity.

Table~\ref{tab:dspi2} shows the descriptive statistics for defect detection rate in programs. It is higher for nametbl and ntree than for cmdline. This result could be due to cmdline having a slightly higher complexity than the other two programs.

\begin{table}[!hbt]
  \caption{Descriptive Statistics for Program Defect Detection Rate in the Replicated Study}
  \label{tab:dspi2}
  \begin{tabular}{lcccc}
    \toprule    
     & & Std. & \multicolumn{2}{c}{95\% Confidence Interval} \\ 
     \cmidrule{4-5}
     Program & Mean & Deviation & Lower Bound & Upper Bound \\    
    \midrule      
     cmdline & 39.315 & 20.407 & 32.700 & 45.931 \\
     nametbl & 53.419 & 14.400 & 48.751 & 58.087 \\
     ntree & 52.564 & 20.065 & 46.059 & 59.068 \\     
    \bottomrule
\end{tabular}
\end{table}

\begin{figure}[!hbt]
\includegraphics[width=4in]{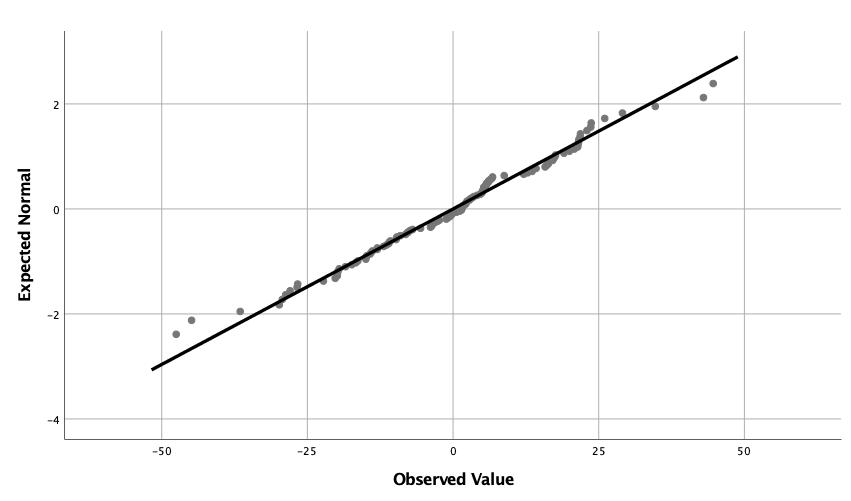}
\caption{Normal Q-Q plot of residuals in the replicated study}
\label{fig:qq_it2}
\end{figure}

We started the analysis of the replicated study data with the same model as in the original study. This time we had neither convergence nor normality issues. The model chosen was: group, program, technique and the program by technique interaction are fixed effects and subject is a random effect. Program is a repeated measures effect, and the covariance structure is Diagonal. Table~\ref{tab:mixed_model_it2} shows the results of the analysis of the best model. Figure~\ref{fig:qq_it2} shows residuals normality. 

\begin{table}[!hbt]
  \caption{Type III Tests of Fixed Effects in the Replicated Study}
  \label{tab:mixed_model_it2}
  \begin{tabular}{lcccc}
    \toprule     

	Source & Numerator df & Denominator df & F & Sig. \\
     \midrule      
 	Intercept & 1 & 42.788 & 858.926 & 0.000 \\
 	Group & 5 & 64.916 & 1.930 & 0.101 \\
 	Technique & 2 & 76.721 & 1.817 & 0.169 \\
 	Program & 2 & 57.386 & 8.308 & 0.001 \\
 	Technique * Program & 4 & 77.253 & 0.796 & 0.532 \\
  
     \bottomrule  
\end{tabular}
\end{table}

Table~\ref{tab:mixed_model_it2} shows that program is statistically significant. Group, technique and the technique by program interaction are not significant. The Bonferroni multiple comparisons tests show that cmdline shows a lower defect detection rate than the other two programs (cmdline$<$(cmdline$=$nametbl)).

Table~\ref{tab:technique_means_it2} and Table~\ref{tab:program_means_it2} show the estimated marginal means for both technique and program.

\begin{table}[!hbt]
  \caption{Estimated Marginal Means for Technique in the Replicated Study}
  \label{tab:technique_means_it2}
  \begin{tabular}{lccccc}
    \toprule     

	 &  & & & \multicolumn{2}{c}{95\% Confidence Interval} \\
	 \cmidrule{5-6}
	Technique & Mean & Std. Error & df & Lower Bound & Upper Bound \\
     \midrule      
 	CR & 52.871 & 2.782 & 106.039 & 47.356 &  58.386 \\
 	BT & 47.041 & 2.765 & 106.927 & 41.560 &  52.523 \\
 	EP & 46.020 & 2.817 & 101.380 & 40.432 &  51.608 \\
 	  
     \bottomrule  
\end{tabular}
\end{table}

\begin{table}[!hbt]
  \caption{Estimated Marginal Means for Program in the Replicated Study}
  \label{tab:program_means_it2}
  \begin{tabular}{lccccc}
    \toprule     

	 &  & & & \multicolumn{2}{c}{95\% Confidence Interval} \\
	 \cmidrule{5-6}
	Program & Mean & Std. Error & df & Lower Bound & Upper Bound \\
     \midrule      
 	cmdline & 39.379  & 3.100 & 36.670 & 33.096 & 45.662 \\
 	nametbl & 53.962 & 2.065 & 38.280 & 49.783 & 58.142 \\
 	ntree & 52.591 & 3.075 & 38.282 & 46.368 & 58.814 \\
  
     \bottomrule  
\end{tabular}
\end{table}

Finally, Figure~\ref{fig:pp_pt_it1} shows the profile plot with error bars for the program by technique interaction. In this case, the profile plot suggests that no interaction exists, which means the non-significance of the interaction in the analysis could not be due to sample size.

\begin{figure}[!hbt]
\includegraphics[width=4.5in]{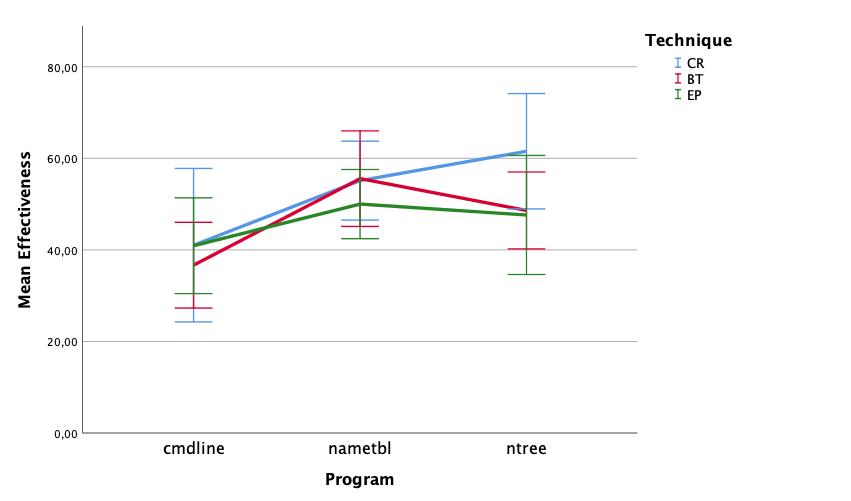}
\caption{Profile plot for program by technique interaction in the replicated study}
\label{fig:pp_pt_it2}
\end{figure}

\section{Joint Analyses}
\label{app:joint_analyses}

\subsection{RQ1.1: Participants’ Perceptions}

Table~\ref{tab:perc_joint} shows the percentage of participants that perceive each technique to be the most effective. We cannot reject the null hypothesis that the frequency distribution of the responses to the questionnaire item (\textit{Using which technique did you detect most defects?}) follows a uniform distribution ($\chi^2$(2,N=60)=1.900, p=0.387). This means that the number of participants perceiving a particular technique as being more effective cannot be considered different for all three techniques. {\bf Our data do not support the conclusion that techniques are differently frequently perceived as being the most effective}.

\begin{table}[!hbt]
  \caption{Participants' Perceptions of Technique Effectiveness in the Joint Analysis}
  \label{tab:perc_joint}
  \begin{tabular}{ccccc}
    \toprule    
     N & CR & BT & EP & Result \\
    \midrule      
    60 & 30.00\% & 28.33\% & 41.67\% & CR=BT=EP \\
    \bottomrule
\end{tabular}
\end{table}

\subsection{RQ1.2: Comparing Perceptions with Reality}

Table~\ref{tab:agreement_joint} shows the value of kappa and its 95\% CI, overall and for each technique separately. We find that all values for kappa with respect to the questionnaire item (\textit{Using which technique did you detect most defects?}) are consistent with lack of agreement, except for CR ($\kappa$$<$0.4, poor). This means that {\bf our data do not support the conclusion that participants correctly perceive the most effective technique for them}.

\begin{table}[!hbt]
  \caption{Agreement between Perceived and Real Technique Effectiveness in the Joint Analysis}
  \label{tab:agreement_joint}
  \begin{tabular}{lccc}
    \toprule    
      & & \multicolumn{2}{c}{95\% Confidence Interval} \\      \cmidrule{3-4}
      & Kappa value & Lower bound & Upper bound \\ \midrule      
 	  Overall & 0.241 & 0.059 & 0.444 \\	
 	  CR & 0.420 & 0.170 & 0.648 \\	
 	  BT & 0.116 & -0.136 & 0.393 \\	
 	  EP & 0.168 & -0.092 & 0.430 \\	

    \bottomrule
\end{tabular}
\end{table}

As lack of agreement cannot be ruled out, we examine whether the perceptions are biased. The results of the Stuart-Maxwell test show that the null hypothesis of existence of marginal homogeneity cannot be rejected ($\chi^2$(2,N=60)=2.423, p=0.298). Additionally, the results of the McNemar-Bowker test show that the null hypothesis of existence of symmetry cannot be rejected ($\chi^2$(3,N=60)=2.552, p=0.466). This means that we cannot conclude that there is directionality when participants' perceptions are wrong. These two results suggest that participants are not differently mistaken about one technique as they are about the others. {\bf Techniques are not differently subject to misperceptions}.

\subsection{RQ1.3: Comparing the Effectiveness of Techniques}

We are going to check if misperceptions could be due to participants detecting the same amount of defects with all three techniques, and therefore being impossible for them to make the right decision. Table~\ref{tab:agreementK_joint} shows the value and 95\% CI of Krippendorff's $\alpha$ overall and for each pair of techniques, for all participants and for every design group (participants that applied the same technique on the same program) separately, and Table~\ref{tab:agreementpK_joint} shows the value and 95\% CI of Krippendorff's $\alpha$ overall and for each program/session. For values with all participants, we can rule out agreement ($\alpha$$<$0.4) except for the case of EP-BT and nametbl-ntree for which the upper bound of the 95\% CIs are consistent with fair to good agreement. However, even in this two cases, 0 belongs to the 95\% CIs, meaning that agreement by chance cannot be ruled out. This means that {\bf participants do not obtain similar effectiveness values when applying the different techniques (testing the different programs) so as to be difficult to discriminate among techniques/programs}. As regards the results for groups, the 95\% CIs are too wide to show reliable results.

\begin{table}[!hbt]
  \caption{Agreement between Percentage of Defects Found with Each Technique in the Joint Analysis}
  \label{tab:agreementK_joint}
  \begin{tabular}{lclccc}
    \toprule    
      & & & Krippendorff's & \multicolumn{2}{c}{95\% Confidence Interval} \\ \cmidrule{5-6}
      Sample & N & &  alpha value & Lower bound & Upper bound \\ \midrule      
 
 	  All & 61 & CR-BT & -0.1074 & -0.4575 & 0.1887 \\	
 	  & & CR-EP & -0.0735 & -0.4893 & 0.2888 \\	
 	  & & EP-BT & 0.2276 & -0.1096 & 0.5574 \\	
 	  & & CR-BT-EP & 0.0264 & -0.1893 & 0.2108 \\	\midrule   
 	  
 	  G1 & 11 & CR-BT & -0.1005 & -0.8416 & 0.5812 \\
 	  & & CR-EP & 0.4410 & 0.0021 & 0.7586 \\
 	  & & EP-BT & -0.1573 & -1.0000 & 0.5615 \\
 	  & & CR-BT-EP & 0.0204 & -0.6032 & 0.4626 \\ \midrule 
 
 	  G2 & 11 & CR-BT & 0.0843 & -0.3277 & 0.4351 \\
 	  & & CR-EP & 0.2121 & -0.1489 & 0.5262 \\
 	  & & EP-BT & 0.4109 & -0.1718 & 0.8474 \\
 	  & & CR-BT-EP & 0.2015 & -0.0429  & 0.4289 \\ \midrule 

	  G3 & 12 & CR-BT & -0.1142 & -0.8008 & 0.4459 \\
 	  & & CR-EP & -0.2242 & -1.0000 & 0.5142 \\
 	  & & EP-BT & 0.2978  &-0.4059  & 0.7313 \\
 	  & & CR-BT-EP & 0.0397 & -0.3692 & 0.3948 \\ \midrule 

	  G4 & 11 & CR-BT & -0.1320 & -0.9523 & 0.4963 \\
 	  & & CR-EP & -0.1295 & -1.0000 & 0.6278  \\
 	  & & EP-BT & 0.7483 & 0.5245 & 0.9161 \\
 	  & & CR-BT-EP & 0.1263 & -0.3618 & 0.5116 \\ \midrule 
 	  
	  G5 & 4 & CR-BT & -0.4382 & -1.0000 & 0.5625 \\
 	  & & CR-EP & -0.2669 & -0.8214 & 0.2538 \\
 	  & & EP-BT & 0.1131 & -0.4175 & 0.6056 \\
 	  & & CR-BT-EP & -0.1471 & -0.5263 & 0.2249 \\ \midrule 

	  G6 & 2 & CR-BT & -0.3766 & -1.0000 & 0.7864 \\
 	  & & CR-EP & -0.5468 & -1.0000 & 0.8683 \\
 	  & & EP-BT & 0.7689 & 0.6768 & 0.9060 \\
 	  & & CR-BT-EP & -0.1187 & -1.0000 & 0.7590 \\
	  
    \bottomrule
\end{tabular}
\end{table}

\begin{table}[!hbt]
  \caption{Agreement between Percentage of Defects Found with Each Program in the Joint Analysis (N=61)}
  \label{tab:agreementpK_joint}
  \begin{tabular}{lccc}
    \toprule    
      & Krippendorff's & \multicolumn{2}{c}{95\% Confidence Interval} \\ \cmidrule{3-4}
      &  alpha value & Lower bound & Upper bound \\ \midrule      
 	  cmdline-nametbl & -0.0942 & -0.5398 & 0.2782 \\	
 	  cmdline-ntree & 0.0551 & -0.2394 & 0.3211 \\	
 	  nametbl-ntree & 0.0685 & -0.3615 &  0.4256 \\	
 	  cmdline-nametbl-ntree & 0.0242 & -0.1831 & 0.2178 \\	
    \bottomrule
\end{tabular}
\end{table}

\subsection{RQ1.4: Cost of Mismatch}

\begin{figure}[!hbt]
\includegraphics[width=3.5in]{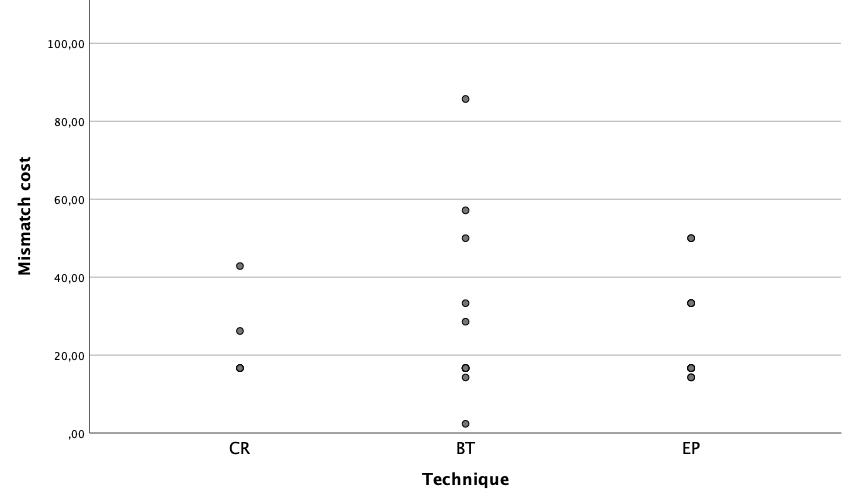}
\caption{Scatterplot for observed mismatch cost in the original study}.
\label{fig:spmc_joint}
\end{figure}

Table~\ref{tab:cost_joint} and Figure~\ref{fig:spmc_joint} show the cost of mismatch. We can see that the CR technique has fewer mismatches compared to the other two. Although the BT and EP techniques have the same number of mismatches, BT shows a higher dispersion. The results of the Kruskal-Wallis test reveal that we cannot reject the null hypothesis of techniques having the same mismatch cost (H(2)=0.034, p=0.983). This means that we cannot claim a difference in mismatch cost between the techniques. The estimated mean mismatch cost is 27pp (median 17pp).

\begin{table}[!hbt]
  \caption{Observed Reduction in Technique Effectiveness for Mismatch. Column 2 shows the number of mismatches out of the total number of participants who perceived the technique as being most effective. Columns 3-5 show the mean, median and standard deviation for mismatch cost (in percentage points)}
  \label{tab:cost_joint}
  \begin{tabular}{ccccc}
    \toprule
     & & \multicolumn{3}{c}{Cost} \\
     \cmidrule{3-5}  
     Technique & No. Mismatches & Mean & Median & Std. Deviation \\
    \midrule     
     CR & 5(18) & 24pp & 17pp & 11 \\
     BT & 12(17) & 30pp & 17pp & 24 \\
     EP & 13(25) & 25pp & 17pp & 13 \\
     \midrule
     TOTAL & 30(60) & 27pp & 17pp & 18 \\
    \bottomrule
\end{tabular}
\end{table}

These results suggest that {\bf the mismatch cost is not negligible} (27pp), {\bf and is not related to the technique perceived as most effective.}

\subsection{RQ1.5: Expected Loss of Effectiveness}

Table~\ref{tab:loss_joint} shows the average loss of effectiveness that should be expected in a project. Again, the results of the Kruskal-Wallis test reveal that we cannot reject the null hypothesis of techniques having the same expected reduction in technique effectiveness for a project (H(2)=5.680, p=0.058). This means we cannot claim a difference in project effectiveness loss between techniques. The mean expected loss in effectiveness in the project is estimated as 13pp.

\begin{table}[!hbt]
  \caption{Observed Reduction in Technique Effectiveness in a Software Project. Column 2 shows the number of (mis)matches. Columns 3-5 show the mean, median and std. deviation for the reduction in effectiveness in the project (in percentage points)}
  \label{tab:loss_joint}
  \begin{tabular}{ccccc}
    \toprule
     & & \multicolumn{3}{c}{Cost} \\
     \cmidrule{3-5}  
     Technique & N & Mean & Median & Std. Deviation \\
    \midrule     
     CR & 18 & 7pp & 0pp & 12 \\
     BT & 17 & 21pp &  17pp & 24 \\
     EP & 25 & 13pp  & 14pp & 16 \\
     \midrule
     TOTAL & 60 & 13pp & 1pp & 18 \\
    \bottomrule
\end{tabular}
\end{table}

These results suggest that {\bf the expected loss in effectiveness in a project is not negligible} (15pp), {\bf and is not related to the technique perceived as most effective.} 

\end{appendices}

\end{document}